\pdfoutput=1
\RequirePackage{ifpdf}
\ifpdf 
\documentclass[pdftex]{sigma}
\else
\documentclass{sigma}
\fi

\numberwithin{equation}{section}

\usepackage{mathrsfs}
\usepackage{multirow}
\usepackage{pdflscape}

\usepackage{tikz-cd}

\newcommand{\graphScale}{2}
\newcommand{\horSep}{1*\graphScale}
\newcommand{\vertSep}{0.75*\graphScale}
\newcommand{\edgeText}{\small}
\newcommand{\nodeText}{\large}
\tikzstyle{bag} = [text width=1em, text centered]
\tikzstyle{end} = [] 

\newcommand{\graphFive}{
\begin{tikzpicture}[sloped]
  \node (a) at (0,0) [bag] {\nodeText $a$};
  \node (b) at (\horSep,0) [bag] {\nodeText $b$};
  \draw [-] (a) to node [above] {\edgeText $0.464 $} (b);
 \end{tikzpicture}
}

\newcommand{\graphSix}{{
\begin{tikzpicture}[sloped]
  \node (a) at (0,0) [bag] {\nodeText $a$};
  \node (b) at (2*\horSep,0) [bag] {\nodeText $b$};
  \node (c) at (\horSep,0) [bag] {\nodeText $c$};
  \draw [-] (a) to node [above] {\edgeText $ 0.739$} (c);
  \draw [-] (b) to node [above] {\edgeText $ 0.6$} (c);
 \end{tikzpicture}
}}

\newcommand{\graphSeven}{
\begin{tikzpicture}[sloped]
  \node (a) at (0,0) [bag] {\nodeText $a$};
  \node (b) at (\horSep,0) [bag] {\nodeText $b$};
  \draw [-] (a) to node [above] {\edgeText $ 0.715$} (b);
 \end{tikzpicture}
}

\newcommand{\graphEight}{{
\begin{tikzpicture}[sloped]
  \node (a) at (0,0) [bag] {\nodeText $a$};
  \node (b) at (1.5\horSep,0) [bag] {\nodeText $b$};
  \node (c) at (0.75\horSep,-0.9*\vertSep) [bag] {\nodeText $c$};
  \draw [-] (a) to node [above] {\edgeText $0.289$} (b);
  \draw [-] (a) to node [below] {\edgeText $0.758$} (c);
  \draw [-] (b) to node [below] {\edgeText $0.293$} (c);
 \end{tikzpicture}
}}

\newcommand{\graphNineReduced}{{
\begin{tikzpicture}[sloped]
  \node (a) at (0,0) [bag] {\nodeText $a$};
  \node (c) at (0,\vertSep) [bag] {\nodeText $b$};
  \node (d) at (\horSep,0) [bag] {\nodeText $c$};
  \node (e) at (-\horSep,0) [bag] {\nodeText $d$};
  \draw [-] (a) to node [above] {\edgeText $0.28$} (c);
  \draw [-] (a) to node [below] {\edgeText $0.308$} (d);
  \draw [-] (a) to node [below] {\edgeText $0.898$} (e);
  \draw [-] (c) to node [above] {\edgeText $0.317$} (d);
  \draw [-] (c) to node [above] {\edgeText $0.928$} (e);
 \end{tikzpicture}
}}

\newcommand{\graphTen}{{
\begin{tikzpicture}[sloped]
  \node (a) at (-\horSep,0) [bag] {\nodeText $a$};
  \node (b) at (\horSep,0) [bag] {\nodeText $b$};
  \node (c) at (\horSep,\vertSep) [bag] {\nodeText $c$};
  \node (d) at (0,\vertSep) [bag] {\nodeText $d$};
   \node (e) at (\horSep+\horSep,\vertSep) [bag] {\nodeText $e$};
   \node (f) at (-\horSep,\vertSep) [bag] {\nodeText $f$};
   \node (g) at (0,0) [bag] {\nodeText $g$};
   \node (h) at (\horSep,\vertSep+\vertSep) [bag] {\nodeText $h$};
  \draw [-] (a) to node [below] {\edgeText $0.754$} (g);
  \draw [-] (g) to node [below] {\edgeText $0.897$} (b);
  \draw [-] (b) to node [below] {\edgeText $0.292$} (d);
  \draw [-] (b) to node [below] {\edgeText $0.997$} (e);
  \draw [-] (c) to node [above] {\edgeText $0.774$} (e);
  \draw [-] (c) to node [above] {\edgeText $0.236$} (d);
  \draw [-] (d) to node [above] {\edgeText $0.290$} (f);
  \draw [-] (d) to node [above] {\edgeText $1.167$} (h);
  \draw [-] (e) to node [above] {\edgeText $0.927$} (h);
 \end{tikzpicture}
}}

\DeclareMathOperator{\sgn}{sgn}
\DeclareMathOperator{\grad}{\rm grad}
\DeclareMathOperator{\Tr}{Tr}

\newcommand{\ignore}[1]{}

\newcommand{\CC}{{\mathscr{C}}}

\newcommand{\R}{{\mathbb{R}}}
\newcommand{\Z}{{\mathbb{Z}}}

\newcommand{\C}{{\mathbb{C}}}
\newcommand{\I}{{\mathbb{I}}}

\newcommand{\CP}{{\mathbb{C}}{{P}}}

\newcommand{\ra}{\rightarrow}

\newcommand{\cd}{\partial}

\newcommand{\wh}{\widehat}

\newcommand{\s}{{\mathsf{s}}}
\newcommand{\ip}[1]{\langle#1\rangle}
\newcommand{\xvec}{{\bf x}}
\newcommand{\vvec}{{\bf v}}
\newcommand{\Xvec}{{\bf X}}
\newcommand{\evec}{{\bf e}}
\newcommand{\qvec}{{\bf q}}
\newcommand{\Avec}{{\bf A}}

\def \d{\mathrm{d}}

\newcommand{\ol}{\overline}

\newcommand{\id}{{\rm Id}}
\newcommand{\vol}{{\rm vol}}

\newcommand{\tauvec}{\mbox{\boldmath{$\tau$}}}

\renewcommand{\phi}{\varphi}

\usepackage{xparse}
\usetikzlibrary{fadings,shapes.arrows,shadows}

\newcommand{\configSize}{0.15\linewidth}

\graphicspath{{./images/}}

\begin{document}
\allowdisplaybreaks

\renewcommand{\thefootnote}{}

\newcommand{\arXivNumber}{2305.18126}

\renewcommand{\PaperNumber}{073}

\FirstPageHeading

\ShortArticleName{Nudged Elastic Bands and Lightly Bound Skyrmions}

\ArticleName{Nudged Elastic Bands and Lightly Bound Skyrmions\footnote{This paper is a~contribution to the Special Issue on Topological Solitons as Particles. The~full collection is available at \href{https://www.emis.de/journals/SIGMA/topological-solitons.html}{https://www.emis.de/journals/SIGMA/topological-solitons.html}}}

\Author{James Martin SPEIGHT~$^{\rm a}$ and Thomas WINYARD~$^{\rm b}$}

\AuthorNameForHeading{J.M.~Speight and T.~Winyard}

\Address{$^{\rm a)}$~School of Mathematics, University of Leeds, Leeds LS2 9JT, UK}
\EmailD{\href{mailto:j.m.speight@leeds.ac.uk}{j.m.speight@leeds.ac.uk}}
\URLaddressD{\url{https://cp1lump.github.io}}

\Address{$^{\rm b)}$~School of Mathematics, University of Edinburgh, Edinburgh, EH9 3FD, UK}
\EmailD{\href{mailto:twinyard@ed.ac.uk}{twinyard@ed.ac.uk}}

\ArticleDates{Received May 30, 2023, in final form September 25, 2023; Published online October 11, 2023}

\Abstract{It has become clear in recent years that the configuration space of the nuclear Skyrme model has, in each topological class, many almost degenerate local energy minima and that the number of such minima grows with the degree (or baryon number) $B$. Rigid body quantization, in which one quantizes motion on the spin-isospin orbit of just one minimum, is thus an ill-justified approximation. Instead, one should identify a (finite-dimensional) moduli space of configurations containing all local minima (for a given $B$) as well as fields interpolating smoothly between them. This paper proposes a systematic computational scheme for generating such a moduli space: one constructs an energy minimizing path between each pair of local minima, then defines the moduli space to be the union of spin-isospin orbits of points on the union of these curves, a principal bundle over a graph. The energy minimizing curves may be constructed in practice using the nudged elastic band method, a standard tool in mathematical chemistry for analyzing reaction paths and computing activation energies. To illustrate, we apply this scheme to the lightly bound Skyrme model in the point particle approximation, constructing the graphs for $5\leq B\leq 10$. We go on to complete the quantization for $B=7$, in which the graph has two vertices and a~single edge. The low-lying quantum states with isospin $1/2$ do not strongly localize around either of the local energy minima (the vertices). Their energies rise monotonically with spin, conflicting with experimental data for Lithium-7.}


\Classification{81T12; 70G45}

\renewcommand{\thefootnote}{\arabic{footnote}}
\setcounter{footnote}{0}

\section{Introduction}

The Skyrme model is a theory of nuclear physics in which nuclei are treated as
topological solitons called Skyrmions~\cite{sky}. Originally proposed in 1962, it was eclipsed as a model of strong interactions by QCD. Interest in the model revived, however, after Witten argued
that it emerges as a low-energy effective theory of QCD in the large $N_c$ (number of colours) limit~\cite{wit-sky2}. In its simplest form, the model has only one dynamical field,
a map $U\colon \R^3\ra {\rm SU}(2)$ from physical space to the Lie group ${\rm SU}(2)\equiv S^3$. After imposition of a suitable boundary condition, this field is classified topologically by an integer-valued invariant $B$, its degree or winding number, which is interpreted
physically as the baryon number of the field configuration. Stable static solutions
of degree $B$ are used to model nuclei of atomic weight $B$, that is, containing $B$ nucleons. Small amplitude travelling waves perturbing the vacuum, $U(x)=\I_2$, are interpreted, after quantization, as $\pi$ mesons.

The Skyrme model provides a coherent description of isolated nucleons (protons and neutrons) but as a model of larger nuclei it faces two (probably linked) challenges. First, it greatly over-estimates the binding energies of nuclei (typically by a factor of 10 or more). Second, in order to make contact with real physics, one must somehow quantize Skyrmion dynamics, and since the model is not renormalizable, there is
inherent ambiguity in this step. The simplest quantization scheme is rigid body quantization: one quantizes motion on the spin-isospin orbit of a single static
solution. This scheme is simple and systematic, and works well for $B$ fairly small and even (it correctly predicts the spin and isospin of the nuclear ground state for
$B=1,2,3,4,6,8$ and $12$ for example~\cite{kru}), but misses important effects in general, particularly for nuclei with small binding energies. This seems to be related to an interesting property of the space of Skyrme fields which has only recently emerged: for a given degree $B$ there are typically very many inequivalent stable static solutions which may be very close in energy and fairly close
in field space. This fact was
first noticed for a variant of the model due to Harland, engineered to have low binding energy, called the lightly bound model~\cite{gilharspe}, but has recently been shown to be generic even in more usual versions of the model~\cite{gudhal}. Rigid body quantization restricts dynamics to the symmetry orbit of just one of these static solutions for each $B$, a radical and ill-justified simplification. What is needed is a quantization scheme in which the field moves on a manifold containing all the low-lying local energy minima as well as fields interpolating between these. Several ingenious schemes of this kind have been proposed~\cite{hal,halkinman,raw-c}, but each is a bespoke construction designed to deal with one particular value of $B$, and their details are often, in part, reverse engineered from the
phenomenology one hopes to reproduce, rather than being derived {\it ab initio} from the Skyrme model.

The current paper presents a first step towards a systematic method for generating an appropriate space of Skyrme configurations, including all known local energy minimizers, and fields smoothly interpolating between them. The idea is that, given a pair of local energy minimizers, one constructs an energy minimizing path in field space between them. That is, to any path joining the two local minima one assigns the maximum energy of all fields along the path; an energy minimizing path is then a path which, among all paths, has the least maximal energy. By construction, the field on this path with maximal energy is a saddle point of energy (of index one). The numerical scheme we will use to construct such a path has the useful and natural property that the two path segments joining the saddle point to each of the local minima are gradient descent curves. By constructing such a path between every pair of distinct local energy minimizers, we generate a graph in field space. Following Rawlinson~\cite{raw}, we propose to quantize motion on the union of symmetry orbits of points on this graph.

The numerical scheme we will use to construct the energy minimizing paths between static solutions is called the {\em nudged elastic band} (NEB) method. It has been quite widely used in mathematical chemistry, but, as far as we are aware, has not previously been applied  in the field of topological solitons. Bessarab et al.\ have used it to construct
annihilation paths from magnetic skyrmions to the vacuum in spin lattice systems, as a means to estimate the stability of magnetic skyrmions to thermal fluctuations
\cite{besuzdjon}. This is conceptually rather different from the problem we address here (since it relies crucially on the spatial discreteness of the system), but the underlying mathematical ideas are identical.

To apply the NEB method to a field theory in $d$ dimensions, one starts with a chain of field configurations roughly equally spaced (in field space) between the two local energy minimizers, then couples neighbouring configurations on the chain with harmonic springs and allows the configurations to move until the energy gradient and spring forces are in equilibrium. Having discretized physical space, this, then, turns into a lattice field theory problem in $(d+1)$ dimensions: the $d$ dimensions of space plus the one dimension along the chain. So, for the Skyrme model, we are faced with a
4-dimensional field theoretic optimization problem, a rather ambitious undertaking for a first foray of the method into soliton dynamics. One could, of course, treat a two-dimensional analogue of the Skyrme model as a warm-up exercise. We have chosen, instead, to apply the method to the lightly bound Skyrme model, where a very effective point particle approximation to Skyrmion dynamics is available~\cite{gilharkirmayspe}.  Unlike in the usual Skyrme model, it is strongly energetically disfavoured for individual Skyrmions to approach one another closely: at modest energies they remain distinct and behave like individual point particles, each carrying an internal orientation which may be
parametrized by a point in ${\rm SU}(2)$. The local energy minimizers of degree $B$ are extremely well approximated by superpositions of $B$ unit Skyrmions positioned on the vertices of a face centred cubic lattice, their internal orientations governed by a colouring rule. So, in this model, there is an obvious point particle approximation to the configuration space, $M_B=\big(\R^3\times {\rm SU}(2)\big)^B$, and dynamics in this space should closely approximate the real field dynamics at low energy.

We will use the NEB method to construct energy minimizing paths between pairs of local energy minimizers in $M_B$ for $B\leq 10$. From this we can determine the energy barriers separating the different static solutions: these are just the energies of the intervening saddle points. We can also construct, for each $B$, the graph motion on which will, after quantization, approximate atomic weight $B$ nuclei. This quantization problem is still highly nontrivial, and the details of its implementation depend
significantly on the topology of the graph and the symmetries of the static solutions represented by its vertices. We illustrate the method by quantizing the $B=7$ sector. In this case, the graph is rather simple: two vertices joined by a single edge, but this suffices to exhibit the main features of the method. In the end, our model predicts a quantum ground state of isospin $1/2$ \big(representing Li${}^7$ or Be${}^7$\big) and spin $1/2$ whose shape is a roughly equal superposition of the two static solutions (an octohedron plus one extra particle, and a~line of 3 particles above a square). This conflicts with nature \big(the ground state of  Li${}^7$ has spin $3/2$\big) and is less successful than Halcrow's vibrational quantization of the $B=7$ Skyrmion in the conventional Skyrme model (which also gets the ground state spin wrong, but gets closer to the low-lying spectrum of excited states)~\cite{hal}. We interpret this as evidence against the lightly bound model rather than the underlying programme of graph quantization being proposed.

\section{The lightly bound Skyrme model}

Assign to any map $U\colon\R^3\ra {\rm SU}(2)$, satisfying the boundary condition $U(\infty)=\I_2$, the energy
\begin{gather}
\label{lbe}
E = \int_{\R^3}
\Bigg[
 (1-\alpha)\left( -\frac12\Tr(R_iR_i) + m^2\Tr(1-U) \right)\nonumber
 \\
\hphantom{E = \int_{\R^3}\bigg[}{}  - \frac{1}{16}\Tr([R_i,R_j][R_i,R_j]) + \alpha\left(\frac12\Tr(1-U)\right)^4 \Bigg] \d^3 x,
\end{gather}
where $R_i=(\cd_i U)U^{-1}$ is the associated right current and $\alpha\in[0,1]$ and
$m>0$ are dimensionless constants. This defines a one-parameter family of (static) Skyrme models, parametrized by $\alpha$, interpolating between the standard Skyrme model with massive pions, at $\alpha=0$, and Harland's unbound model, at $\alpha=1$. The latter has a topological lower energy bound $E\geq 8\pi B$, attained only for $B=1$~\cite{har1}.
As usual, the degree of $U$ is
\[
B= -\frac{1}{24\pi^2}\int_{\R^3}\epsilon_{ijk}\Tr(R_iR_jR_k) \, \d^3 x.
\]
Hence, Skyrmions in this model are marginally unbound: they can approach the energy
$8\pi B$ arbitrarily closely, by diverging to spatial infinity, but they can never
attain this energy. So, at $\alpha=1$, the model has zero binding energy (binding energy being, by definition, $B$ times the energy of a unit Skyrmion minus the infimum of $E$ over the charge $B$ sector) while at
$\alpha=0$ it has unphysically large binding energy. Harland's idea was that, somewhere in between there should be a model in which binding energies are comparable to those of real nuclei. Setting the pion mass parameter $m=1$, as is usual, one finds that this does indeed happen, for $\alpha\approx 0.95$~\cite{gilharspe}. We set $m=1$ and $\alpha=0.95$ for the remainder of this paper, and refer to this variant of the Skyrme model as the lightly bound model.

Although this is, in its mathematical formulation, the least radical of all the
so-called near BPS models~\cite{adasanwer2,sut-holographic}, it turns out that its Skyrmions, that is, local minimizers of $E$, are markedly different from those of the usual ($\alpha=0$) model. They consist of $B$ quite distinct almost spherically symmetric charge 1 subunits, arranged on compact subsets of the vertices of a face centred cubic lattice, internally oriented so as to be in an attractive channel with each of their nearest neighbours~\cite{gilharkirmayspe, gilharspe}.
The internal orientations can be understood as follows. Decompose the FCC lattice
\[
\big\{(x_1,x_2,x_3)\in\Z^3\colon x_1+x_2+x_3\cong 0\bmod 2\big\}
\]
into 4 simple cubic sublattices labelled by $([x_1],[x_2])\in \Z_2\times\Z_2$ and think of these 4 values as colours. Assign to these 4 colours $c$ the ${\rm SU}(2)$ matrices $L_c$
\[
([0],[0])\mapsto \I_2,\qquad
([0],[1])\mapsto -{\rm i}\tau_1,\qquad
([1],[0])\mapsto -{\rm i}\tau_2,\qquad
([1],[1])\mapsto -{\rm i}\tau_3,
\]
where $\tau_i$ are the usual Pauli matrices. Then a unit Skyrmion placed at a vertex
$\vvec$ with colour $c$ is the image of the standardly oriented unit Skyrmion $U_H$ (the $H$ stands for ``hedgehog'') under isorotation by $L_c$, that is,
\[
U(\xvec)=L_cU_H(\xvec-\vvec)L_c^{-1}.
\]
By construction, this places all nearest neighbours in the lattice in the attractive channel.

All local minima found by lattice field simulations\footnote{With one exception: at $B=8$ the 5th lowest energy minimizer, while still exhibiting $8$ distinct unit Skyrmions, is a bound state of two tetrahedra whose constituents lie on two different FCC lattices.} are approximately of this form and, furthermore, given any reasonably compact order $B$ subset of the FCC lattice, one finds a local energy minimizer close to this, with internal orientations (approximately) given by the colouring rule. Hence, as $B$ grows, the number of local energy minimizers grows very rapidly~\cite{gilharkirmayspe}.

It is natural, therefore, to develop a point-particle approximation to the model. By careful study of two Skyrmion scattering processes, it was determined that the interaction energy of two unit Skyrmions placed at positions $\xvec_1$ and $\xvec_2$, with internal orientations
$q_1,q_2\in {\rm SU}(2)$ (relative to $U_H$) is well approximated by
\begin{gather*}
V_{int}(\xvec_1,q_1,\xvec_2,q_2) = V_0(|\mathbf{X}|) + V_1(|\mathbf{X}|)\Tr(R(Q)) + V_2(|\mathbf{X}|) \frac{\mathbf{X}\cdot R(Q)\mathbf{X}}{|\mathbf{X}|^2},
\end{gather*}
where $\Xvec=\xvec_1-\xvec_2$, $Q=q_1^{-1}q_2$ and $V_0$, $V_1$, $V_2$ are certain explicit functions of $|\Xvec|$, all exponentially small at large $|\Xvec|$ whose precise form
is given in~\cite{gilharkirmayspe}. Here, and henceforth, we use $R(q)$ to denote the
${\rm SO}(3)$ matrix whose action on $\R^3$ coincides with the adjoint action of $q\in {\rm SU}(2)$ on the Lie algebra $\mathfrak{su}(2)$ under the standard identification
\[
\R^3\ra \mathfrak{su}(2),\qquad \xvec\mapsto -\frac{{\rm i}}{2}\xvec\cdot\tauvec.
\]
That is,
\[
q\left(-\frac{{\rm i}}{2}\xvec\cdot\tauvec\right)q^{-1}=:-\frac{{\rm i}}{2}(R(q)\xvec)\cdot\tauvec.
\]
We may now approximate charge $B$ static Skyrmions by finding local minima of the
total point-particle energy \smash{$V_{\rm pp}\colon \big(\R^3\times {\rm SU}(2)\big)^B\ra\R$,}
\begin{gather}\label{ppe}
V_{\rm pp}((\xvec_1,q_1),\ldots,(\xvec_B,q_B))=BM+\sum_{i=1}^{B-1}\sum_{j=i+1}^B V_{int}(\xvec_i,q_i,\xvec_j,q_j),
\end{gather}
the sum of the rest energy of $B$ unit Skyrmions and their pairwise interaction energies. The results are in excellent agreement with lattice field simulations for the range of $B$ in which the latter have been comprehensively prosecuted ($1\leq B\leq 8$). This gives us confidence to discard the original field theoretic model~\eqref{lbe} and replace it by the much simpler point particle system~\eqref{ppe}. We make the model dynamical by ascribing a rest mass $M$ and moment of inertia~$L$ to each point Skyrmion, obtaining the Lagrangian
\begin{gather}\label{paux}
L=\frac12 \sum_{a=1}^B\big(M|\dot{\xvec}_a|^2+ L|\dot{q}_a|^2\big)-V_{\rm pp},
\end{gather}
where, in computing the length of $\dot{q}_a$, we have given ${\rm SU}(2)$ the unit sphere metric (with respect to which, it is crucial to note, $-{\rm i}\tau_i$ are orthonormal, rather than $-{\rm i}\tau_i/2$).
The constants $M$ and~$L$ are obtained by fitting to lattice field theory simulations,
and are found to be $M=93.09$, $L=217.20$.\footnote{There is an error in the computation of $L$ in~\cite{gilharkirmayspe}, resulting in a value which is 4 times too small. This is due to an inconsistency in identifications of $\mathfrak{su}(2)$ with $\R^3$. Luckily, the inconsistency cancels itself out, and the results on rigid body quantization presented in~\cite{gilharkirmayspe} are correct.}

In summary, the numerical problem we are faced with is, for a given value $B$, first, to~find
all local minima of $V_{\rm pp}\colon M_B\ra\R$, where $M_B=\big(\R^3\times {\rm SU}(2)\big)^B$, and second, for each distinct pair of local minima, to construct an energy minimizing path between them. Here we must face an important subtlety: points in $M_B$ which differ by a permutation of the particle labels, and/or sign-flips of (some of) the internal orientations actually represent precisely the same field configuration in the original Skyrme model, and should not be considered distinct. This is because the field obtained by superposing $B$ oriented hedgehog configurations is clearly invariant under changes in how we label the hedgehog centres, and $R(-q)=R(q)$, so the internal orientation of each charge one subunit is invariant under $q\mapsto-q$. To formulate this precisely, let $\mathfrak{S}_B$ denote the permutation group on $B$ objects and $\Z_2=\{1,-1\}$ with multiplication. On the cartesian product $P_B:=\mathfrak{S}_B\times \Z_2^B$
define the group multiplication law
\[
(\sigma,\s)\cdot(\sigma',\s')=
\big(\sigma\circ\sigma',
 \big(s_1,s'_{\sigma^{-1}(1)},\ldots,\s_B\s'_{\sigma^{-1}(B)}\big)\big).
\]
We call $P_B$ the group of perm-flips: it has a natural right action on
$M_B$ by permuting the particle labels and flipping the orientations:
\[
(\sigma,\s)\colon \ ((\xvec_1,q_1),\ldots,(\xvec_B,q_B))\mapsto
\big(\big(\xvec_{\sigma(1)},\s_{\sigma(1)}q_{\sigma(1)}\big),\ldots,
\big(\xvec_{\sigma(B)},\s_{\sigma(B)}q_{\sigma(B)}\big)\big).
\]
Then points in $M_B$ lying in the same $P_B$ orbit represent identical fields, and should be identified.

Local minimizers of~\eqref{ppe} were studied in~\cite{gilharkirmayspe}, which proposed that they closely resemble size $B$ subsets of a face centred cubic lattice.  This paper used a zero-temperature annealing algorithm to locate the minima, which is not accurate or efficient enough for our purposes (see the next section).
Hence we have recomputed all local minima using a gradient flow algorithm using the results of~\cite{gilharkirmayspe} as initial data. The results are qualitatively unchanged, but the minimal energies are reduced, typically in the 3rd (or 2nd, for large $B$) decimal place. A comparison of these results with those of~\cite{gilharkirmayspe} for $B=3,4,\ldots,10$ can be found in  Appendix \ref{appendix1}. We continue the convention of labelling local minima with their particle number $B$ and a lower-case Latin label~$a,b,c,\ldots$, so that alphabetical order reproduces their energetic order (so the lowest charge~$7$ solution is labelled $7a$, the next lowest $7b$ etc.).
A comparison between the energy minimizers found in the point particle model and those obtained by full numerics on the field theory was also made in
\cite{gilharkirmayspe} for $1\leq B\leq 8$, with good qualitative and
quantitative agreement found.

\section{Nudged elastic band method}

Given two local minimizers of $V_{\rm pp}$ with the same particle number $B$, labelled $B\alpha$ and $B\beta$ say,  we seek an energy minimizing path (EMP) between them. To find said EMP, we employ the nudged elastic band (NEB) method~\cite{besuzdjon}. This entails constructing a curve $\Gamma\colon[0,1]\ra M_B$ with $\Gamma(0)=B\alpha$ and
$\Gamma(1)=B\beta$ such that
\[
\max_{s\in[0,1]} V_{\rm pp}(\Gamma(s))
\]
is as small as possible.  To describe the method, it is helpful to assume at first that the configuration space in which the curve lives, $M_B$, is a finite-dimensional vector space. We will then describe the modifications required to accommodate the nonlinear nature of $M_B$ in our application.

So,
consider first the analogous problem for some potential function $V\colon M\ra\R$ and a pair of local minima $\alpha,\beta\in M=\R^k$. To find an EMP connecting
$\alpha$ to $\beta$, we first discretize the curve $\Gamma$, replacing it by
an ordered set of points $\alpha=v_1,v_2,\ldots, v_n=\beta$.
  We then treat each configuration $v_i$ as a notional particle in $M$, coupled to its neighbours with notional springs, of some uniform spring constant $\kappa$, subject to a force due to the potential $V$ and forces due to the coupling springs.  Crucially,  we project the force due to the potential $V$ {\em orthogonal} to the path, and the force due to the springs {\em parallel} to the path.  The second projection ensures that an EMP is a gradient descent curve, while the first projection evenly distributes the points $v_i$ along the EMP. This is most easily seen after taking the continuum limit, to which we will turn shortly. So, having chosen a unit vector $\hat\tau_i$ approximating the unit tangent to the curve at the point $v_i$, the total force experienced by the $i$-th particle is
\[
F_i^{\rm tot}=- \grad V(v_i)|_\perp+F_i^{\rm spring}\big|_{\parallel},
\]
where
\begin{gather*}
u|_\parallel = \langle u, \hat\tau_i \rangle \hat\tau_i,\qquad
u|_\perp = u- u|_\parallel,\qquad
F_i^{\rm spring} = -\kappa (v_{i+1} - 2v_i+v_{i-1}),
\end{gather*}
and $\ip{\cdot,\cdot}$ denotes the Euclidean inner product on $M$.
The start and end points of our chain are fixed, $v_1=\alpha$, $v_n=\beta$, but we allow all the intermediate points $v_i(t)$, $i=2,3,\ldots,n-1$, to move under {\em arrested} Newton flow. That is, we solve the coupled system of
ODEs
\[
\ddot{v_i}=F_i^{\rm tot}
\]
starting from some initial configuration with $\dot{v}(0)=0$, but subject to the arresting criterion that if, at any time,
\[
\big\langle\dot{v},F^{\rm tot}\big\rangle=\sum_{i=2}^{n-1}\big\langle\dot{v}_i,F^{\rm tot}_i\big\rangle<0,
\]
we set $\dot{v}=0$ and restart the flow. This flow efficiently relaxes the chain to an equilibrium point, at which $F^{\rm tot}_i=0$. In practice, we solve the flow using a 4th order Runge--Kutta scheme with fixed time step, and declare the chain to be at equilibrium when the sup norm of $F^{\rm tot}$ falls below some prescribed tolerance.

An appealing feature of this algorithm is that the chains it produces are
discrete approximants of unions of gradient descent curves for the potential $V$, parametrized with constant speed, from some saddle point $v_*$ between $v_1$ and $v_n$. To see this, imagine taking the continuum limit of the equilibrium chain, that is, taking both the spring constant $\kappa$ and the number of particles in the chain $n$ to infinity such that $\kappa/n=1$, and taking $v_i=v(i/n)$ for some continuous curve $v(s)$. Then $F^{\rm spring}_i$ approaches the limit
\[
F^{\rm spring}(s)=-\frac{{\rm d}^2 v}{{\rm d}s^2}
\]
and $\hat\tau_i$ the limit
\[
\hat\tau(s)=\frac{{\rm d}v}{{\rm d}s}\bigg/\left|\frac{{\rm d}v}{{\rm d}s}\right|.
\]
Since the chain is at equilibrium, $F^{\rm tot}=0$, and, resolving this condition into its perpendicular and parallel components, one concludes that
\begin{gather}
- \grad V(v(s))|_\perp=0,\label{kao}\\
\frac{{\rm d}^2 v}{{\rm d}s^2}\bigg|_{\parallel}=0\label{ish}.
\end{gather}
Condition~\eqref{kao} implies that $\grad V(v(s))$ is parallel to the curve at
$v(s)$, and hence that $v$ is a gradient descent curve (or ascent curve, depending on how one orients it). Note that this condition actually results from the projection of $F^{\rm spring}$ parallel to $\hat\tau$. Condition~\eqref{ish} implies that $v(s)$ has constant speed and results from the projection of $\grad V$ perpendicular to $\hat\tau$. The point on the curve $v_*$ of maximal energy is a critical point of $V$ ($\grad V|_\perp=0$ as for any point on the curve, and $\grad V|_\parallel=0$ since $V$ is maximal along the curve) with exactly one unstable direction, parallel to the curve, and hence an index $1$ saddle point of $V$.

The configuration space for the application we have in mind, $M_B=\big(\R^3\times S^3\big)^B$ is not a~vector space. However, it has a canonical isometric embedding into $M=\R^{7B}$, obtained via the standard embedding of $S^3$ into $\R^4$.
So we construct a chain of points in $M$ between $\alpha$ and~$\beta$, each of which lies on the submanifold $M_B$, and then apply the algorithm described above to this chain, but with all forces and velocities projected tangent to $M_B$ (orthogonally, using the ambient metric on $M$), and, after each time step, each point in the chain projected onto $M_B$ by radial projection
$\R^4\ra S^3$ in each factor.

A key computational choice in the NEB method is the construction of the unit vector $\hat\tau_i$ approximating the tangent vector to the curve at
$v_i$~\cite{besuzdjon}.   The simplest is using the normalized
displacement between neighbouring configurations,
\[
\hat\tau_i = \frac{v_{i+1} - v_{i-1}}{|v_{i+1} - v_{i-1}|}.
\]
This approximation can cause issues when the curvature is large as the lattice sites will not be evenly spaced along the curve $\Gamma$.  A better choice is to average between the line segments of both neighbouring configurations,
\[
\tau_i = \frac{v_i - v_{i-1}}{|v_i - v_{i-1}|} + \frac{v_{i+1} - v_{i}}{|v_{i+1} - v_{i}|},
\]
which is then normalised to give $\hat\tau_i$. This is better at maintaining uniform spacing between neighbours on the chain.

Finally, both the previous approaches have issues if the energy changes too rapidly along the path. This can be ameliorated by using a piecewise definition for $\hat\tau_i$ chosen so that, for a given point $v_i$ we take the normalised displacement between $v_i$ and the neighbour with higher energy,
\[
\tau = \begin{cases}\tau^+ & 	\text{if }  E_{i+1} > E_i >E_{i-1},\\ \tau^- & \text{if }  E_{i+1} < E_i < E_{i-1}, \end{cases}
\]
where
$
\tau^+ = v_{i+1} - v_i$,  $\tau^-  = v_i - v_{i-1}$.
Note that if the chosen configuration corresponds to a local minimum or maximum on $\Gamma_{\alpha\beta}$, then the tangent becomes
\[
\tau_i = \begin{cases} \tau_i^+ \Delta E_i^\text{max} + \tau_i^- \Delta E_i^\text{min} & \text{if} \ E_{i+1} > E_{i-1}, \\
\tau_i^+ \Delta E_i^\text{min} + \tau_i^- \Delta E_i^\text{max} & \text{if} \ E_{i+1} < E_{i-1},
\end{cases}
\]
where
\begin{gather*}
\Delta E^{\text{max}}_i  = \max( |E_{i+1} - E_i|, |E_{i-1} - E_i|),\qquad \Delta E^{\text{min}}_i = \min( |E_{i+1} - E_i|, |E_{i-1} - E_i|).
\end{gather*}
Finally, the tangent vector is normalized.

  For our simulations, we tried all three methods of choosing the tangent vector and for most paths the choice made little difference in the outcome. This is in part due to the large number of lattice sites used to discretize the paths.  We did find that the final method proved more robust in the few cases where the {EMP} transitioned via many other local minima.  Note that if this approach is expanded to the full Skyrme model, then the computational cost would necessitate a far coarser discretization of the path and the final method presented here may compensate for this loss of accuracy.

It is also worth noting that the magnitude of $\kappa$ will affect the convergence speed for the algorithm. If $\kappa$ is too large then the optimal time step ${\rm d}t$ (while ensuring stability) is dominated by the spring force along the path, while the gradient orthogonal to the path relaxes slowly.  Alternatively, if $\kappa$ is too small, then the spring force relaxes slowly.  When running the algorithm, we used trial and error to determine a workable value, finding that $\kappa=10$ works acceptably across the range of results presented.

To construct a sensible initial condition we now need to carefully reconsider the isometries of the model.  As each local minimizer is only unique up to these isometries, we must choose the optimal position, spatial orientation, isospace orientation, labelling and sign flips for the two fixed boundary configurations.  This is not an issue for the continuous symmetries, as there are zero potential paths that the NEB method can use to compensate for a poor choice. However, changing one of the two discrete symmetries,  produces a curve with a non-zero energy barrier.

First, without loss of generality,  we fix the labelling and choice of sign for the configuration on the left boundary ($s=0$). For the right boundary ($s=1$), we assume that the ``optimal'' choice of isometries is the one that minimises the distance $|v_1 - v_n|$ between the boundary configurations.  Hence, we loop through the possible labellings and spin flips, and for each choice minimize over the other isometries. We then select the choice of isometries that provides the shortest configuration distance.

Even if we assume that a given {EMP} exists, this approach will not always work.  If we choose a~poor labelling or sign choice, then the resulting path will feature multiple saddlepoints and more than 2 minima, where the additional minima will correspond to applying a discrete symmetry to one of the boundary configurations, effectively changing the poor boundary condition.  Hence,  our algorithm upon completion checks all local minima not on the boundaries. Each configuration is used as an initial condition and a gradient flow algorithm was applied to minimise~$V_{\rm pp}$.  If the resulting local minimum of this gradient flow simulation matches either boundary configuration up to symmetries, we can simply use this as an improved boundary condition for a new simulation.

The method presented above still has one significant issue:  the number of equivalent discrete configurations scales factorially with $B$ and quickly becomes impractical for large $B$.  Hence, it is more practical to guess at a reasonable labelling and allow the algorithm to correct the additional minima that occur until a good boundary condition is found.  Of course there is no guarantee that another labelling would not provide an alternative, longer, but lower energy path to the one found, but checking all labellings is computationally impossible.

It is interesting to note that the problem of permutation equivalence is absent if we apply the NEB method to the full field theory: both the start and end configurations are simply (lattice approximants to) maps $\R^3\ra {\rm SU}(2)$, with no labelling assumed. While the full field theory is, in this sense, simpler, extending the NEB method to it will be a numerically intensive task. Each field configuration can be approximated on a regular $N^3$ grid, with spacing $h>0$.  Hence, the discretized configuration space would be the manifold $M_\text{Skyrme} = ({\rm SU}(2))^{N^3} \subset \mathbb{R}^{4N^3}$.  The algorithm then follows similarly to above,  where each EMP is approximated by a chain of configurations $v_i \in \mathbb{R}^{4N^3}$. To resolve the structure of a skyrmion of modest charge, say $B=7$, usually requires
$N\sim 200$, so the NEB method must construct a path in a space of dimension
$4N^3\sim 3.2\times 10^7$, which compares with $7B=49$ for the point particle model. Clearly this is immensely more computationally costly, and it will be necessary to use a much coarser discretization of the path itself (i.e., use fewer notional particles $v_i$ in our chain). If we are to apply the NEB method to more conventional versions of the Skyrme model, which do not have a good point particle approximation, there is no obvious alternative to paying
this cost.

\subsection{Results}

\begin{landscape}
\begin{figure}\centering
\begin{tabular}{ccccccccc}
\includegraphics[width=\configSize, trim={0.2cm 0.0cm 0.25cm 0.1cm},  clip]{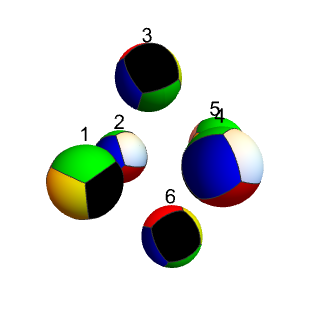} & & \includegraphics[width=\configSize, trim={0.2cm 0.0cm 0.25cm 0.1cm},  clip]{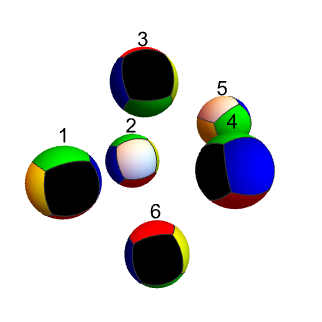}&
& \includegraphics[width=\configSize, trim={0.2cm 0.0cm 0.25cm 0.1cm},  clip]{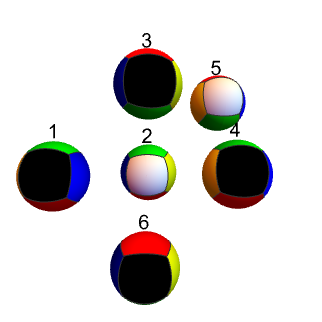} &  & \includegraphics[width=\configSize, trim={0.2cm 0.0cm 0.25cm 0.1cm},  clip]{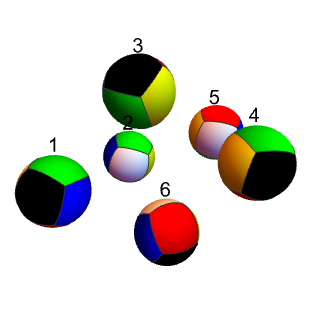}&  & \includegraphics[width=\configSize, trim={0.2cm 0.0cm 0.25cm 0.1cm},  clip]{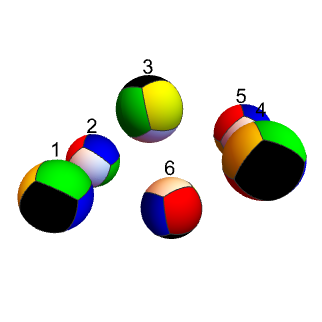}\\
{\Large \bf{a}} &{ \large $\boldsymbol{\rightarrow}$}& {\large saddle point} & { \large $\boldsymbol{\rightarrow}$} & {\Large \bf{c}} & { \large $\boldsymbol{\rightarrow}$} & {\large saddle point} & { \large $\boldsymbol{\rightarrow}$}& {\Large \bf{b}}\\
\includegraphics[width=\configSize, trim={0.3cm 0.5cm 0.4cm 0.2cm},  clip]{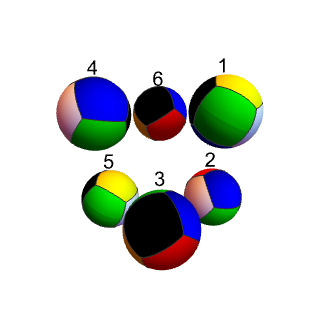} & & \includegraphics[width=\configSize, trim={0.3cm 0.5cm 0.4cm 0.2cm},  clip]{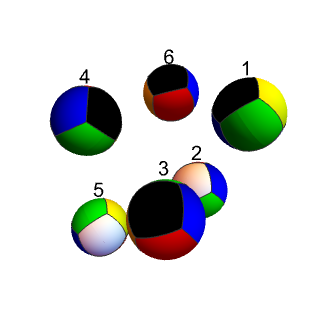}&
& \includegraphics[width=\configSize, trim={0.3cm 0.5cm 0.4cm 0.2cm},  clip]{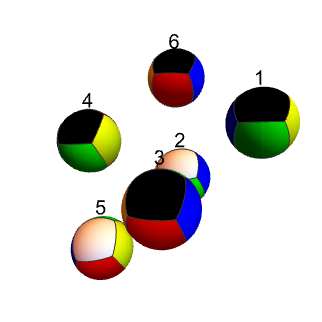} &  & \includegraphics[width=\configSize, trim={0.3cm 0.5cm 0.4cm 0.2cm},  clip]{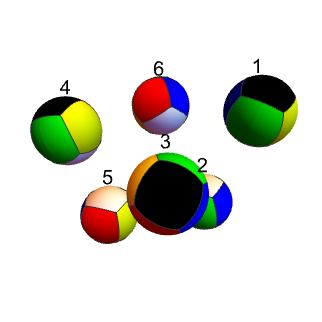}&  & \includegraphics[width=\configSize, trim={0.3cm 0.5cm 0.4cm 0.2cm},  clip]{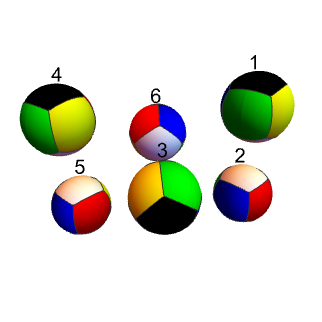}\\
\end{tabular}
\caption{A plot of the $B=6$ transition $\Gamma_{ab}$ from configuration $6a$ to $6b$ via the local minimizer $6c$. Note that as the graph for $B=6$ in Table~\ref{tab:graphs} is the line $a-c-b$ this transition describes the entire graph.  Each point particle is represented by a sphere centred on its position $x_i$ and oriented according to $q_i$. Each sphere has been coloured by patches of colour, with white/black on the top/bottom and blue,green,red and yellow going clockwise around the equator (the so-called juggler's ball colouring, see~\cite{gilharspe} for a more detailed discussion). Both the top and bottom row show the same configurations but from different angles (the perspective is constant along each row).  We have plotted the highest energy configuration between each local minimizer, which corresponds to a~saddle point of the energy.}
\label{fig:B6configs}
\end{figure}
\end{landscape}

\begin{figure}[t]\centering
\includegraphics[width=0.5\linewidth]{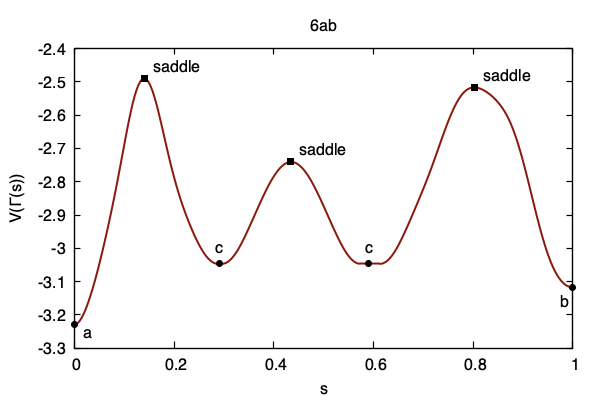}\includegraphics[width=0.5\linewidth]{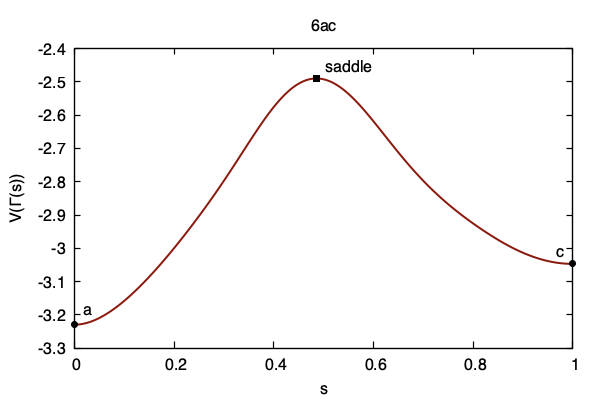}
\includegraphics[width=0.5\linewidth]{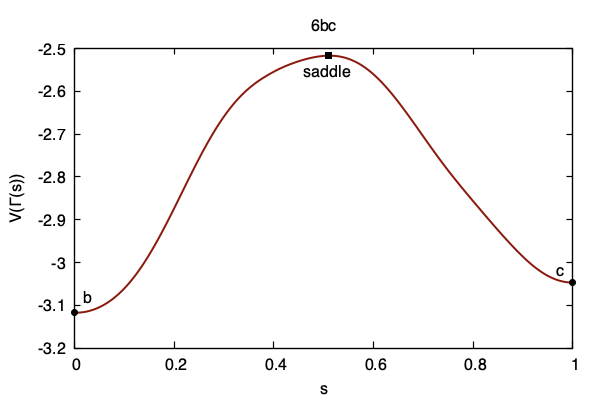}

\caption{Plots of the potential energies $V_{\rm pp}$ of the energy minimizing paths for $B=6$ found using NEB, with saddle points and minima labelled.  Note that the first path transitions from configuration $6a$ to $6b$ via the local minimum $6c$, while the other paths are direct. $6ab$ is the union of $6ac$, $6bc$ reversed and an additional $6cc$ curve that constitutes a relabelling and is artificial. This can be seen in the graph for $B=6$ in Table~\ref{tab:graphs}, while the configurations at each critical point are plotted in Figure~\ref{fig:B6configs}.}
\label{fig:B6paths}
\end{figure}

We used the above method for $B \leq 10$, with a typical number of configurations in a path $n=500$, obtaining the results displayed in Table~\ref{tab:graphs}.  Here $\Gamma_{\alpha\beta}$ is the EMP from $\alpha$ to $\beta$ and we recall that the labelling convention is that alphabetical order reproduces ascending energy order.  The final column displays the reduced configuration spaces in the form of a graph for each degree $B$.  If a vertex connecting two nodes of the graph is not included, it means a minimal energy path between the minimizers was found, but it had additional local minima on the path that correspond to other local minimizers. These were identified using a similar algorithm as explained for identifying improved boundary conditions above, checking each local minimum and comparing with the list of possible local minima.  This is best seen in the $B=6$ graph, where to transition from $a$ to $b$, we must surprisingly go via the highest energy local minimum~$c$ (and two intervening saddle points).  This results in the energy path seen in Figure~\ref{fig:B6paths}, where we can see the first plot contains multiple minima and saddle points.  The path goes from $a$ via a saddle point to $c$, then it performs a relabelling of the configuration, then transitions to $b$.
A~sequence of snapshots along the path $6ab$ can be seen in Figure~\ref{fig:B6configs}. Since the EMP from $6a$ to $6b$ travels via $6c$, the graph for $B=6$ is a line, this constitutes the entire graph. Plots of the energy along the direct paths $6ac$ and $6bc$ are also shown in Figure~\ref{fig:B6paths} for comparison. By contrast, for $B=8$ we also have a graph of 3 vertices, but there is a direct path between each pair, so the graph forms a loop rather than a line.

It is worth noting that when a path $\Gamma_{\alpha\beta}$ contains multiple local minima, if they do not correspond to the list of known minima we have discovered a new local minimiser.  For example, this is the case for the configurations $10g$ and $10h$.  In addition, as the degree increases the number of local minima increases significantly (note many of these minima will not be included in Appendix \ref{appendix1} as an exhaustive search has not been made for large $B$).  This means the graph becomes increasingly complicated and the paths become increasingly difficult to find numerically.

\begin{table}[t!]\centering
\begin{tabular}{|@{\,}c@{\,}|@{\,}c@{\,}|@{\,}c@{\,}c@{\,}c@{\,}|@{\,}c@{\,}|}
\hline
B&Path & Initial energy & Final energy & Saddle energy & Graph\\
\hline
\multirow{2}{*}{5} & \multirow{2}{*}{$\Gamma_{ab}$} & \multirow{2}{*}{$-2.338$} & \multirow{2}{*}{$-2.184$} & \multirow{2}{*}{$-1.874$} & \multirow{2}{*}{\graphFive} \\
 & & & & & \\
\hline
\multirow{2}{*}{6} & $\Gamma_{ac}$ & $-3.229$ & $-3.047$ & $-2.490$ & \multirow{2}{*}{\graphSix}\\
& $\Gamma_{bc}$ & $-3.117$ & $-3.047$ & $-2.517$& \\
\hline
\multirow{2}{*}{7} & \multirow{2}{*}{$\Gamma_{ab}$} & \multirow{2}{*}{$-4.057$} & \multirow{2}{*}{$-3.895$} & \multirow{2}{*}{$-3.342$} & \multirow{2}{*}{\graphSeven} \\
 & & & & & \\
    \hline
\multirow{4}{*}{8} & $\Gamma_{ab}$ & $-4.889$ & $-4.869$ & $-4.600$ & \multirow{4}{*}{\graphEight}\\
    & $\Gamma_{ac}$ & $-4.889$ & $-4.781$ & $-4.131$ & \\
    & $\Gamma_{bc}$ & $-4.869$ & $-4.781$ & $-4.576$ & \\
    & & & & & \\
    \hline
\multirow{5}{*}{9} & $\Gamma_{ab}$ & $-5.664$ & $-5.598$ & $-5.385$ & \multirow{5}{*}{\graphNineReduced} \\
 & $\Gamma_{ac}$ & $-5.664$ & $-5.483$ & $-5.357$ & \\
 & $\Gamma_{ad}$ & $-5.664$ & $-5.460$ & $-4.766$ & \\
 & $\Gamma_{bc}$ & $-5.598$ & $-5.483$  & $-5.281$ & \\
 & $\Gamma_{bd}$ & $-5.598$ & $-5.460$ & $-4.670$ & \\
 \hline
 \multirow{9}{*}{10} & $\Gamma_{ag}$ & $-6.443$ & $-6.133$ & $-5.358$ & \multirow{9}{*}{\graphTen} \\
 & $\Gamma_{bd}$ & $-6.442$ & $-6.284$ & $-6.071$ & \\
 & $\Gamma_{be}$ & $-6.442$ & $-6.277$ & $-5.445$ & \\
 & $\Gamma_{bg}$ & $-6.442$ & $-6.133$ & $-5.546$ & \\
 & $\Gamma_{cd}$ & $-6.307$ & $-6.284$ & $-6.071$ & \\
 & $\Gamma_{ce}$ & $-6.307$ & $-6.277$ & $-5.534$ & \\
 & $\Gamma_{df}$ & $-6.284$ & $-6.194$ & $-5.994$ & \\
 & $\Gamma_{dh}$ & $-6.284$ & $-6.121$ & $-5.118$ & \\
 & $\Gamma_{eh}$ & $-6.277$ & $-6.121$ & $-5.350$ & \\
 \hline
\end{tabular}
\caption{Table containing each unique energy minimizing path (EMP) between local energy minimizers found for $B \leq 10$. The paths are oriented so that they start at the lower minimizer. Column 4 records the maximum energy along the curve which is, by construction, the energy of an index 1 saddle point.
The rightmost column is the resulting graph representing the reduced configuration space. Each edge represents and EMP and is labelled with the energy barrier that must be surmounted to move from the lower energy vertex to the higher (that is, the energy of the intervening saddle point minus the energy of the lower vertex).  Note that all energies quoted are interaction energies: they do not include the total Skyrmion rest mass $BM$.}\label{tab:graphs}
\end{table}

If we consider this as a test case for the full Skyrme model,  it has been shown that the number of local minima does increase as the degree increases~\cite{gudhal}, similar to the point particle model. However, the number of minima that will occur in the EMPs will be far less, as the discrete re-labelling symmetry is absent.

\begin{figure}[t]\centering
\begin{tabular}{ccccc}
\includegraphics[width=0.25\linewidth, trim={2cm 3.0cm 3.2cm 1.1cm},  clip]{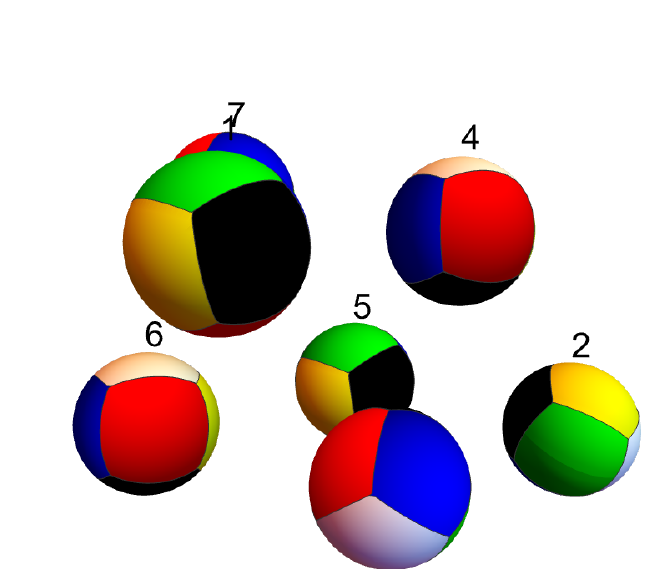} & & \includegraphics[width=0.25\linewidth, trim={2cm 3.0cm 3.2cm 1.1cm},  clip]{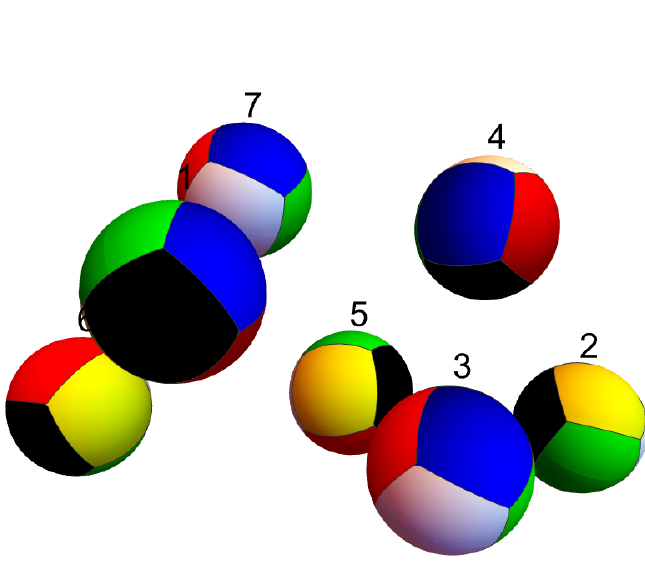}&
& \includegraphics[width=0.25\linewidth, trim={2cm 3.0cm 3.2cm 1.1cm},  clip]{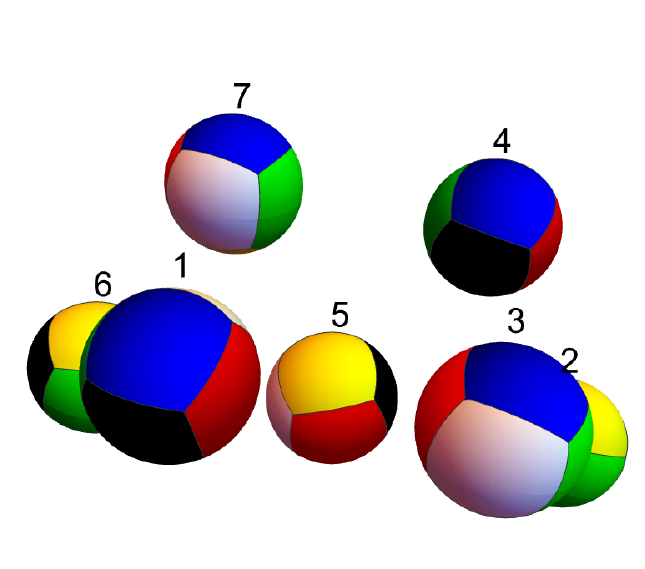}\\
{\Large \bf{a}} &{ \large $\boldsymbol{\rightarrow}$}& {\large saddle point} & { \large $\boldsymbol{\rightarrow}$} & {\Large \bf{b}} \\
\includegraphics[width=0.25\linewidth, trim={2.3cm 0.5cm 2.7cm 1cm},  clip]{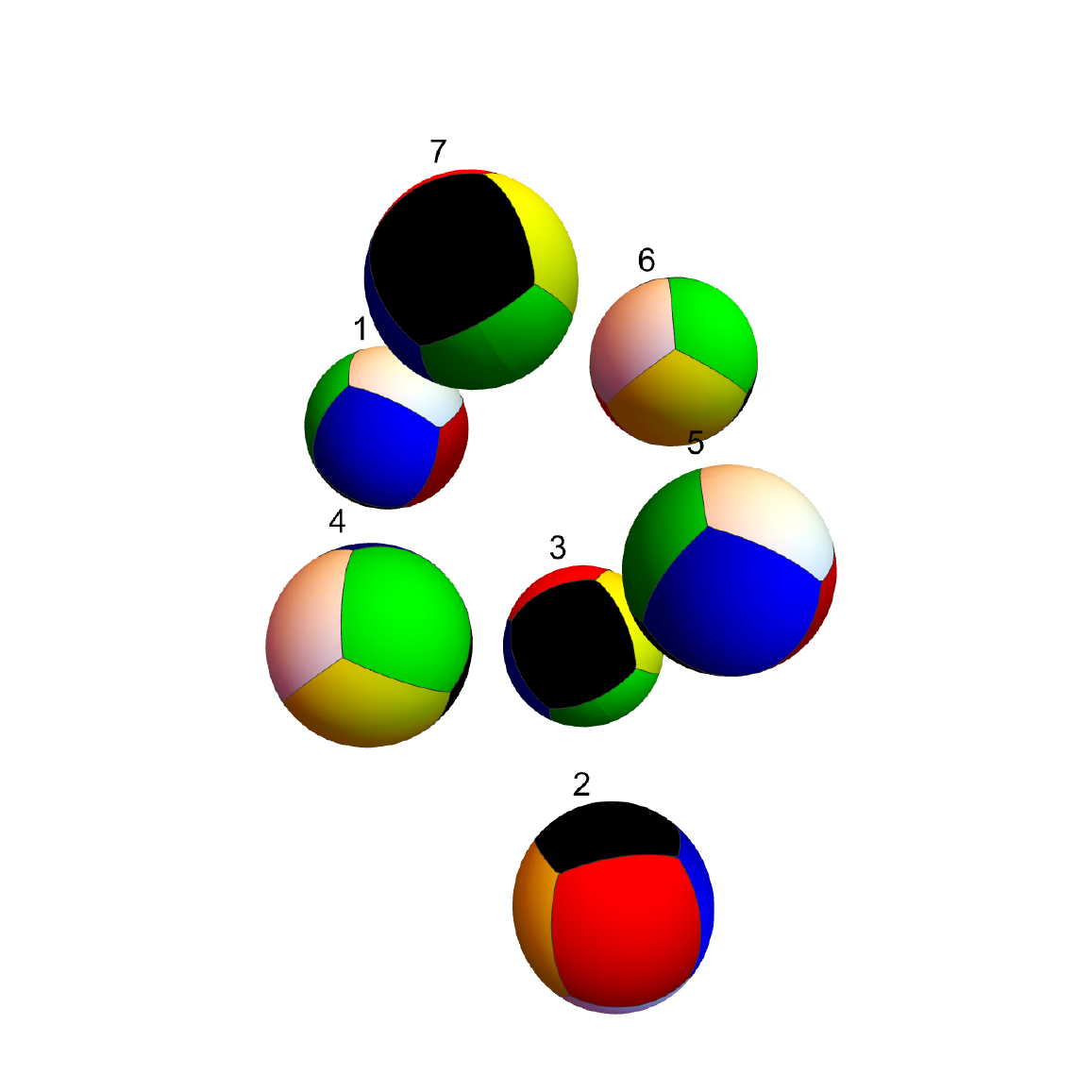} & & \includegraphics[width=0.25\linewidth, trim={2.3cm 0.5cm 2.7cm 1cm},,  clip]{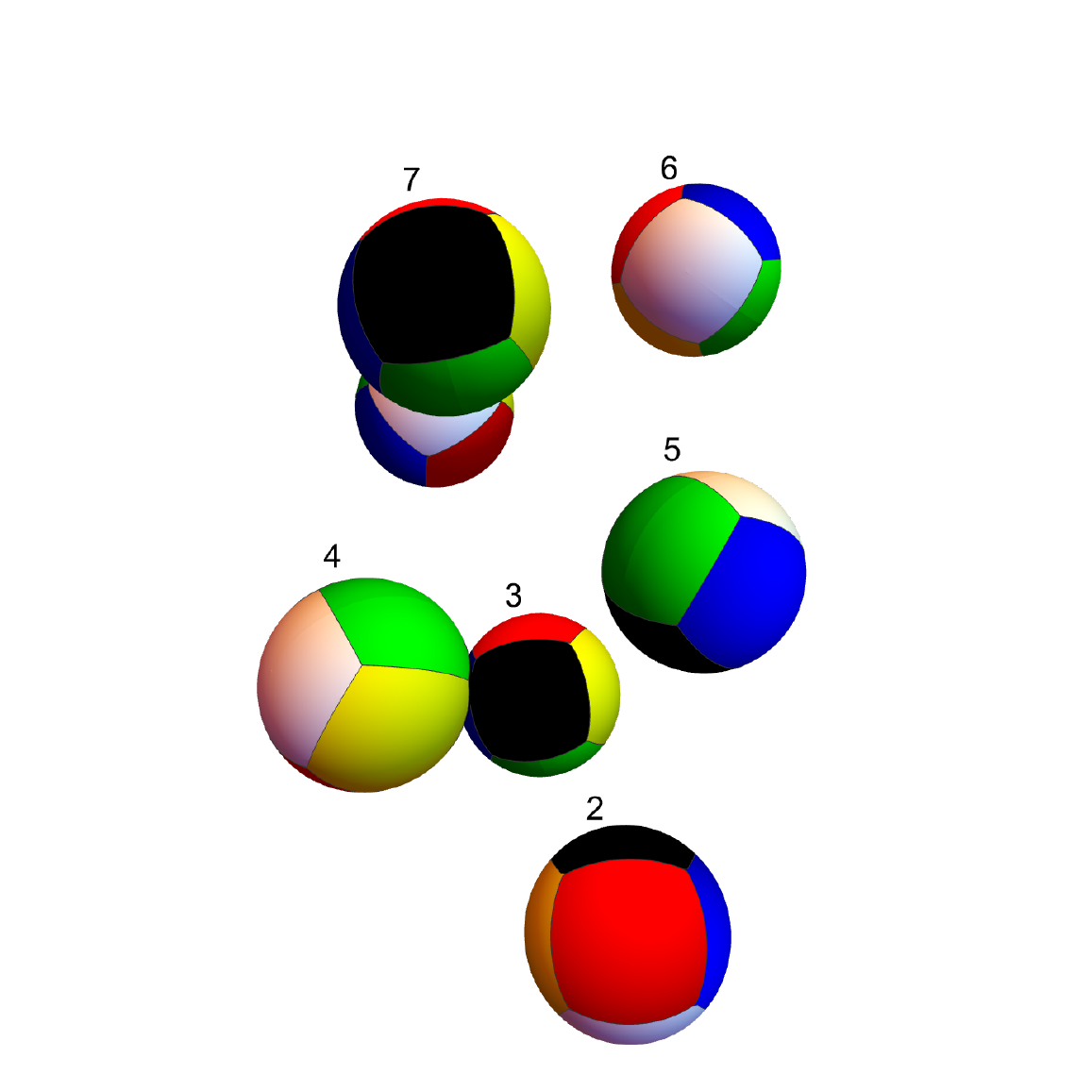}&
& \includegraphics[width=0.25\linewidth, trim={2.3cm 0.5cm 2.7cm 1cm},  clip]{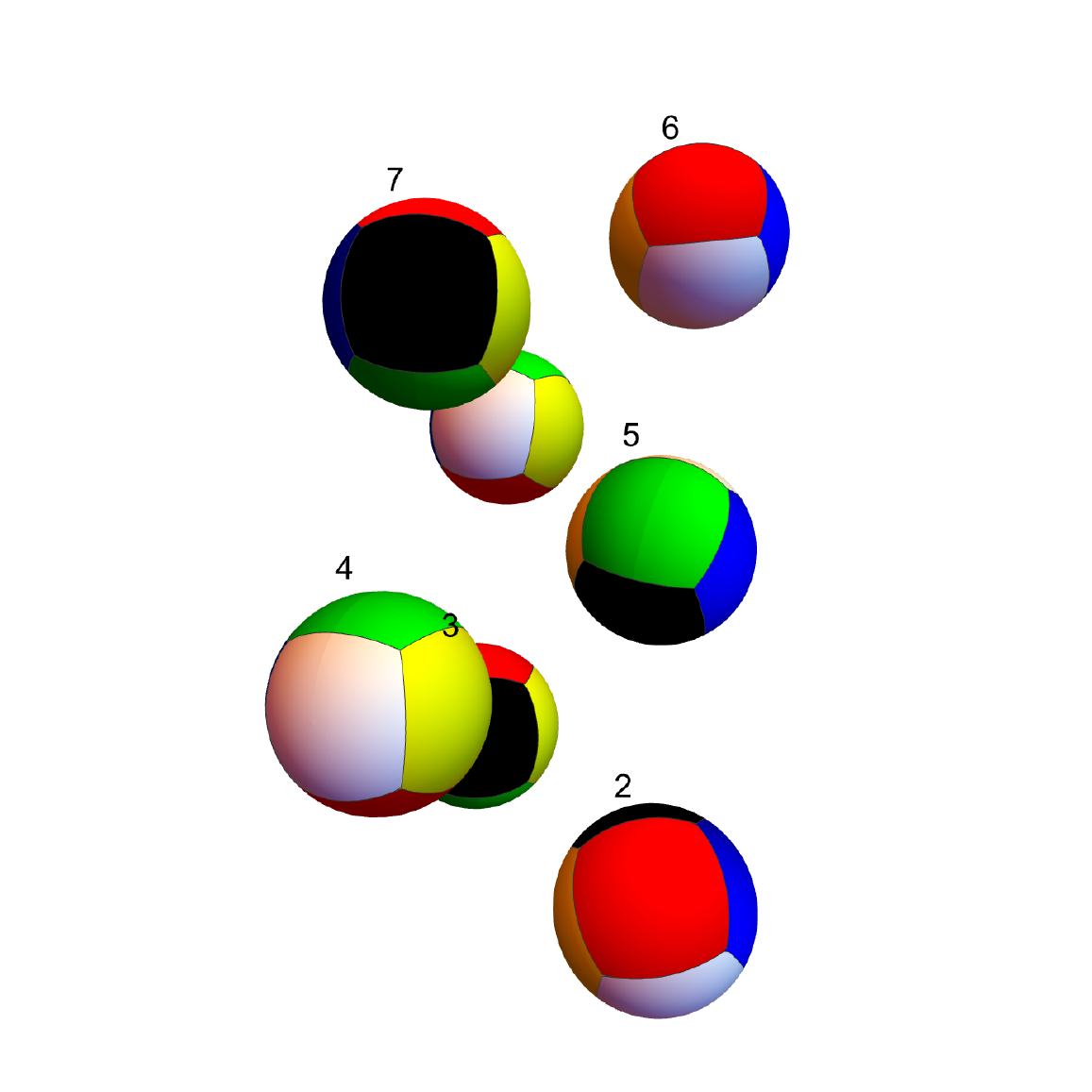}\\
\end{tabular}
\caption{A plot of the $B=7$ transition $\Gamma_{ab}$ from configuration $7a$ to $7b$, with conventions as in Figure~\ref{fig:B6configs}, seen from two different viewpoints (top row versus bottom row). Note that as the graph for $B=7$ in Table~\ref{tab:graphs} is the line $a-b$ this transition describes the entire graph.}
\label{fig:B7configs}
\end{figure}

\section{Quantization}

To quantize the dynamics of $B$ lightly bound Skyrmions, we introduce a wavefunction
\[
\psi\colon \ M_B\ra\C.
\]
Note that this does {\em not} descend to a function $M_B/P_B\ra\C$ since nucleons are fermions. Rather, we demand that, for all $x\in M_B$ and permflips $(\sigma,s)\in P_B$,
\begin{gather}\label{frc}
\psi((\sigma,\s)\cdot x)=(-1)^{\sgn(\sigma,\s)}\psi(x),
\end{gather}
where the sign of a permflip is
$\sgn(\sigma,\s):=\sgn(\sigma)\s_1\s_2\cdots \s_B$. This antisymmetry property of~$\psi$ can be derived from the Finkelstein--Rubinstein constraints applied to the original field theory~\cite{gilharkirmayspe}.\looseness=1

In principle, the dynamics of the system is determined by the natural Hamiltonian operator on $L^2(M_B)$ associated with the particle Lagrangian~\eqref{paux}. This system is too complicated as it stands however, so we make a further approximation. Having identified a graph of energy-minimizing curves $\Gamma$ inside $M_B$, we assume that $\psi$ vanishes outside $G\Gamma$, the union of $G$-orbits of points on $\Gamma$, where
$G={\rm SU}(2)\times {\rm SU}(2)$ is the isospin-spin group. This acts on $M_B$ on the left by
\[
(g,h)\colon \ (\xvec_a,q_a)\mapsto \big(R(h)\xvec_a,hq_ag^{-1}\big).
\]
So we assume the wavefunction is localized only on configurations which, up to isospin-spin symmetry, lie on $\Gamma$. Let us denote the edges of $\Gamma, \Gamma_1,\Gamma_2,\ldots,\Gamma_n$, each parametrized by a coordinate $s\in[0,1]$, and denote by $G\Gamma_j$ the union of $G$-orbits of points on the edge $\Gamma_j$. For $B>2$, the isotropy group of every point on $\Gamma$ is trivial, so we may identify $G\Gamma_j\equiv [0,1]\times G$. The wavefunction reduces, therefore, to a collection of $n$
functions $\psi_j\colon [0,1]\times G\ra\C$, one for each edge, satisfying boundary conditions determined by the graph $\Gamma$, and the symmetries of the vertex configurations, descending from the Finkelstein--Rubinstein constraint~\eqref{frc}. We will return to the issue of boundary conditions shortly.

\subsection[The metric on G Gamma\_j]{The metric on $\boldsymbol{G\Gamma_j}$}

The kinetic energy defined by the particle Lagrangian~\eqref{paux} equips $M_B$ with a Riemannian metric
\[
g_{M_B}=\sum_{a=1}^B\big(M|\d\xvec_a|^2+L|\d q_a|^2\big),
\]
which induces a metric $g_{G\Gamma_j}$ on $G\Gamma_j\cong [0,1]\times G$.
The kinetic term in the Hamiltonian operator acting on $\psi_j$ is the Laplace--Beltrami operator on $[0,1]\times G$ defined by this metric. So our first task is to compute this metric.

By definition, $g_{G\Gamma_j}$ is invariant under the natural left action of $G$ on $[0,1]\times G$, so is uniquely determined by the one-parameter family of inner products it defines on $T_{(s,(\I_3,\I_3))}[0,1]\times G$. This vector space is spanned by $\cd/\cd s, \theta_1,\theta_2,\ldots,\theta_6$ where
\[
\theta_i=-\frac{{\rm i}}{2}\tau_i\oplus 0,\qquad
\theta_{i+3}=0\oplus-\frac{{\rm i}}{2}\tau_i,
\]
so $\theta_{1,2,3}$ generate isorotations while
$\theta_{4,5,6}$ generate rotations. If $\Gamma_j$ is the curve
$\{(\xvec_a(s),q_a(s))\colon 0\leq s\leq 1\}$, then the $G$ part of the metric
may be interpreted as the isospin-spin inertia tensor of the configuration $\Gamma_j(s)$. This was computed in~\cite{gilharkirmayspe}. Relative to the basis $\{\theta_1,\ldots,\theta_6\}$ it takes the form
\begin{gather}\label{sgf}
\big(g_{G\Gamma_j}(\theta_i,\theta_k)\big)=
\Lambda(s)=M\sum_{a=1}^B\left(\begin{matrix}0&0\\
0&|\xvec_a|^2\I_3-\xvec_a\xvec_a^{\mathsf{T}}\end{matrix}\right)
+\frac{L}{4}\sum_{a=1}^B\left(\begin{matrix}\I_3&-R(q_a)\\ -R(q_a)^{\mathsf{T}}&\I_3
\end{matrix}\right).
\end{gather}
The basis vector $\cd/\cd s$ has squared length
\[
g_{G\Gamma_j}(\cd/\cd s, \cd/\cd s)=
g_\CC(\{\dot{\xvec}_a,\dot{q}_a\},\{\dot{\xvec}_a,\dot{q}_a\})
=M\sum_{a=1}^B|\dot{\xvec}_a(s)|^2+L\sum_{a=1}^B|\dot{q}_a(s)|^2,
\]
which must be constructed numerically from the curve.

It remains to compute $g_{G\Gamma_j}(\cd/\cd s, \theta_j)$. For $i=1,2,3$, $\theta_i$ generates the tangent vector
\[
\bigg\{\bigg(0,q_a\frac{{\rm i}}{2}\tau_i\bigg)\bigg\},
\]
whose inner product with $\cd/\cd s$ is
\begin{align*}
g_{G\Gamma_j}(\cd/\cd s, \theta_i)
&=g_{\CC}\bigg(\{\dot{\xvec}_a,\dot{q}_a\},\bigg\{\bigg(0,q_a\frac{{\rm i}}{2}\tau_i\bigg)\bigg\}\bigg)=L\sum_{a=1}^B\bigg\langle q_a^{-1}\dot{q}_a,\frac{{\rm i}}{2}\tau_i\bigg\rangle_{\mathfrak{su}(2)}\\
&=\frac{L}{2}\sum_{a=1}^B\left(q_{a,0}\dot{q}_{a,i}-\dot{q}_{a,0}q_{a,i}-
\epsilon_{ijk}q_{a,j}\dot q_{a,k}\right),
\end{align*}
where we have identified $q_a=q_{a,0}\I_2-{\rm i} q_{a,j}\tau_j$.
For $i=1,2,3$, $\theta_{i+3}$ generates the tangent vector
\[
\bigg\{\bigg(\evec_i\times\xvec_a,-\frac{{\rm i}}{2}\tau_iq_a\bigg)\bigg\},
\]
whose inner product with $\cd/\cd s$ is
\begin{align*}
g_{G\Gamma_j}(\cd/\cd s, \theta_{i+3})
&=g_{\CC}\bigg(\{\dot{\xvec}_a,\dot{q}_a\},\bigg\{\bigg(\evec_i\times \xvec_a,-\frac{{\rm i}}{2}\tau_iq_a\bigg)\bigg\}\bigg)\\
&=M\sum_{a=1}^B\evec_i\cdot(\xvec_a\times \dot{\xvec}_a)
+
L\sum_{a=1}^B\bigg\langle\dot{q}_aq_a^{-1},-\frac{{\rm i}}{2}\tau_i\bigg\rangle_{\mathfrak{su}(2)}\\
&=M\sum_{a=1}^B\evec_i\cdot(\xvec_a\times \dot{\xvec}_a)
+
\frac{L}{2}\sum_{a=1}^B\left(-q_{a,0}\dot{q}_{a,i}+\dot{q}_{a,0}q_{a,i}-
\epsilon_{ijk}q_{a,j}\dot q_{a,k}\right).
\end{align*}

In summary, the metric (with respect to the basis $\cd_s,\theta_1,\ldots,\theta_6$) takes the form
\[
g_{G\Gamma_j}=\left(\begin{matrix}
f(s) & \begin{array}{cc} \Avec^I(s) & \Avec^J(s) \end{array} \\
\begin{array}{c} \Avec^I(s) \\ \Avec^J(s) \end{array} & \Lambda(s)
\end{matrix}\right),
\]
where
\begin{gather*}
f(s) =  M\sum_{a=1}^B|\dot{\xvec}_a(s)|^2+L\sum_{a=1}^B|\dot{q}_a(s)|^2,\\
\Avec^I(s) = \frac{L}{2}\sum_{a=1}^B\left(q_{a,0}(s)\dot\qvec_{a}(s)-\dot{q}_{a,0}(s)\qvec_a(s)
-\qvec_a(s)\times\dot\qvec_a(s)\right),\\
\Avec^J(s) = M\sum_{a=1}^B\xvec_a(s)\times\dot\xvec_a(s)-\Bigg(\Avec^I(s)+L\sum_{a=1}^B\qvec_a(s)\times
\dot\qvec_a(s)\Bigg),
\end{gather*}
and $\Lambda(s)$ is defined in~\eqref{sgf}.

\subsection{Computing the Laplacian}

To compute the Laplacian, it is convenient to follow Rawlinson~\cite{raw} and reexpress the metric in the form
\[
g_{G\Gamma_j}=\left(\begin{matrix}
q(s)+B^{\mathsf{T}}(s)\Lambda(s)B(s) & B^{\mathsf{T}}(s)\Lambda(s)\\
\Lambda(s)B(s) & \Lambda(s)
\end{matrix}\right),
\]
where
\begin{gather*}
A(s) = \left(\begin{matrix}\Avec^I\\ \Avec^J\end{matrix}\right),\qquad
B(s) = \Lambda^{-1}(s)A(s),\qquad
q(s) = f(s)-A^{\mathsf{T}}(s)\big(\Lambda^{-1}(s)\big)^{\mathsf{T}}A(s).
\end{gather*}
Then
\[
g_{G\Gamma_j}=\left(\begin{matrix} 1 & B^{\mathsf{T}} \\ 0 & \I_6 \end{matrix}\right)
\left(\begin{matrix} q & 0 \\ 0 & \Lambda \end{matrix}\right)
\left(\begin{matrix} 1 & 0 \\ B & \I_6\end{matrix}\right),
\]
from which we deduce that
\[
g^{-1}=
\left(\begin{matrix}
\dfrac{1}{q} & -\dfrac{B^{\mathsf{T}}}{q}\vspace{1mm}\\
-\dfrac{B}{q} & \Lambda^{-1}+\dfrac{BB^{\mathsf{T}}}{q}\end{matrix}\right)
\]
and
\[
|g|=\sqrt{|\Lambda|q}.
\]

Let us define $\sigma_1,\ldots,\sigma_6$ to be the left invariant one forms on $G$ dual to the vector fields $\theta_i$. Note that
\begin{gather}\label{ar}
\d\sigma_1=-\sigma_2\wedge\sigma_3,\qquad
\d \sigma_4=-\sigma_5\wedge\sigma_6
\end{gather}
and cyclic perms of these (perms of $\{1,2,3\}$ and $\{4,5,6\}$). This follows from the Maurer--Cartan equation for $G$.

Consider a general smooth function $\psi\colon [0,1]\times G\ra \R$. Then
\[
\d\psi=\sum_i\theta_i(\psi)\sigma_i+\psi_s \d s,
\]
so
\[
|\d\psi|^2=\sum_{i,j}\bigg(\Lambda^{-1}_{ij}+\frac{B_iB_j}{q}\bigg)\theta_i(\psi)\theta_j(\psi)-2\sum_i \frac{B_i}{q}\theta_i(\psi)\psi_s+\frac{1}{q}\psi_s^2.
\]
The volume form on $[0,1]\times G$ is
\[
\vol=\sqrt{|\Lambda(s)|q(s)}\, \d s\wedge\sigma_1\wedge\cdots\wedge\sigma_6.
\]
Hence
\begin{gather*}
\|\d\psi\|_{L^2}^2 = \int_{[0,1]\times G}\d s\wedge \sigma_{1\dots6}\sqrt{|\Lambda|q}
\\
\hphantom{\|\d\psi\|_{L^2}^2 = }{} \times \bigg\{ \sum_{i,j}\bigg( \Lambda^{-1}_{ij}+\frac{B_iB_j}{q}\bigg)\theta_i(\psi)\theta_j(\psi)-2 \sum_i  \frac{B_i}{q}\theta_i(\psi)\psi_s+\frac{1}{q}\psi_s^2\bigg\}
=: T_0+T_1+T_2,
\end{gather*}
where $T_\alpha$ denotes the collection of terms containing $\alpha$
$s$-derivatives.
Now
\[
T_0=\int \d s\wedge \sigma_1\wedge\cdots\wedge\sigma_6\bigg\{
\sum_{i}\theta_i(\lambda_i)
-\sqrt{|\Lambda|q}\psi\sum_{i,j}\left(\Lambda^{-1}_{ij}+\frac{B_iB_j}{q}\right)\theta_i\theta_j(\psi)\bigg\},
\]
where
\[
\lambda_i:=\sum_j\sqrt{|\Lambda|q}\left(\Lambda^{-1}_{ij}+\frac{B_iB_j}{q}\right)\psi
\theta_j(\psi)
\]
is a collection of 6 smooth functions on $[0,1]\times G$. We claim that for any such function, and any $i\in\{1,2,\ldots,6\}$,
\begin{gather}\label{arre}
\int_{[0,1]\times G}\d s\wedge \sigma_1\wedge\cdots\wedge\sigma_6\theta_i(\lambda)=0.
\end{gather}
To see this, let us define $\Sigma_i=\sigma_1\wedge\cdots \wedge\wh{\sigma_i}\wedge
\cdots\wedge \sigma_6$, where the hat denotes an omitted form. So $\Sigma_1=\sigma_2\wedge \sigma_3\wedge\cdots\wedge \sigma_6$ etc. Note that every one of these 5 forms is closed (in fact, exact) by~\eqref{ar}. Then
\begin{align*}
\d s\wedge \sigma_1\wedge\cdots\wedge\sigma_6\theta_i(\lambda)
&=(-1)^{i+1}\d s\wedge \theta_i(\lambda)\sigma_i\wedge\Sigma_i\nonumber\\
&=(-1)^{i+1}\d\lambda_i\wedge\Sigma_i\wedge\d s =\d\big((-1)^{i+1}\lambda_i\Sigma_i\wedge\d s\big)
\end{align*}
since $\Sigma_i\wedge \d s$ is closed. Hence, by Stokes's theorem
\[
\int_{[0,1]\times G}\d s\wedge \sigma_1\wedge\cdots\wedge\sigma_6\theta_i(\lambda)=
\int_{\{1\}\times G}(-1)^{i+1}\lambda \,\d s\wedge \Sigma_i-\int_{\{0\}\times G}(-1)^{i+1}\lambda\,\d s\wedge\Sigma_i.
\]
But both these integrals vanish since the restriction of the integrand to any submanifold
 $\{s\}\times G$ is identically zero.

 Hence
 \[
T_0=\int_{[0,1]\times G}\vol\: \psi\biggl(-\sum_{i,j}\bigg(\Lambda^{-1}_{ij}+\frac{B_iB_j}{f}\bigg)\theta_i\theta_j\biggr)\psi.
\]

Similarly,
\begin{align*}
T_1&=-2\int_{[0,1]\times G}\d s\wedge \sigma_1\wedge\cdots\wedge\sigma_6 \sum_{i}\theta_i(\lambda_i)
+2\int_{[0,1]\times G}\vol\: \psi\sum_{i}\frac{B_i}{q}\theta_i(\psi_s)\\
&=\int_{[0,1]\times G}\vol\: \psi\bigg(2\sum_i \frac{B_i}{f}\theta_i\frac{\cd\:}{\cd s}\bigg)\psi,
\end{align*}
where
\[
\lambda_i:=\sqrt{\frac{|\Lambda|}{q}}B_i\psi_s\psi
\]
and we have again used~\eqref{arre}.

Finally,
\begin{align*}
T_2={}&\int_{[0,1]\times G}\d s \wedge\sigma_{1\cdots 6}
\left\{
\frac{\cd\: }{\cd s}
\left(
\sqrt{\frac{|\Lambda|}{q}}\psi\psi_s
\right)
-\psi\frac{\cd\: }{\cd s}
\left(
\sqrt{\frac{|\Lambda|}{q}}\psi_s
\right)
\right\} \nonumber \\
={}&\int_{[0,1]\times G}\d
\left(
	\sqrt{\frac{|\Lambda|}{q}}\psi\psi_s\sigma_{1\cdots 6}
\right)
-\int_{[0,1]\times G}\vol\: \psi
\left(
	\frac{1}{\sqrt{|\Lambda|q}}\frac{\cd \: }{\cd s}
	\left(
		\sqrt{\frac{|\Lambda|}{q}}\frac{\cd \: }{\cd s}
	\right)
\right)\psi \nonumber\\
={}&
\sqrt{\frac{|\Lambda(1)|}{q(1)}}\int_G \psi(1)\psi_s(1)
-\sqrt{\frac{|\Lambda(0)|}{q(0)}}\int_G \psi(0)\psi_s(0) \nonumber \\
&{}+\int_{[0,1]\times G}\vol\:\psi\left(-\frac{1}{\sqrt{|\Lambda|q}}\frac{\cd \: }{\cd s}\left(
\sqrt{\frac{|\Lambda|}{q}}\frac{\cd \: }{\cd s}\right)\right)\psi.
\end{align*}

By identifying
\[
\|\d\psi\|_{L^2}^2=\ip{\psi,\Delta\psi}_{L^2},
\]
we deduce that
\begin{gather}\label{Lap}
\Delta=-\sum_{i,j}
\left(
	\Lambda^{-1}_{ij}+\frac{B_iB_j}{q}
\right)
\theta_i\theta_j
+2\sum_{i}\frac{B_i}{q}\theta_i\frac{\cd \: }{\cd s}
-\frac{1}{\sqrt{|\Lambda|q}}\frac{\cd \: }{\cd s}\left(\sqrt{\frac{|\Lambda|}{q}}\frac{\cd\: }{\cd s}\right).
\end{gather}
Further, to ensure that $\Delta$ is self-adjoint, we should restrict it to functions for which
\[
\int_{G\times\{s\}}\psi(s)\psi_s(s)=0
\]
at $s=0$ and $s=1$.

The complete Hamiltonian, as it acts on the component of the wavefunction on $G\Gamma_j$, is
\begin{gather}\label{Ham}
H=\frac{\hbar^2}{2}\Delta+V(s),
\end{gather}
where $V(s)$ is the restriction to $\Gamma_j$ of the point-particle potential $V_{\rm pp}$ defined in~\eqref{ppe}.

We have gone through the derivation of $\Delta$ in some detail, rather than simply transcribing the formula for the Hamiltonian
quoted
in~\cite{raw}, since the latter formula appears to be mistaken. In particular, the Hamiltonian  in~\cite{raw} does not appear to be self adjoint (with respect to the natural $L^2$ inner product defined by $g$)  unless $\Lambda$ is constant.

\subsection{Isospin, spin and the FR constraints}

Recall that the wavefunction must satisfy the identity~\eqref{frc}. On $G\Gamma_j$, this identity produces a~constraint if and only if the $G$-orbit of $x(s)$ contains images of $x(s)$ under permflips. Since $G$ is the symmetry group, this happens precisely when $x(s)$ has nontrivial symmetries.

Note that {\em every} configuration is symmetric under both $(g_0,h_0)=(-\I_2,\I_2)$ and
$(g_0,h_0)=(\I_2,-\I_2)$, the compensating permflip being (in both cases)
\[
(\sigma,\s)=(\id,-1,-1,-1,\ldots,-1),
\]
whose sign is $(-1)^B$. So $\psi$ on $G\Gamma_j$ must be even (odd) under both antipodal maps
\[
(s,g,h)\mapsto(s,-g,h),\qquad (s,g,h)\mapsto (s,g,-h)
\]
if $B$ is even (odd).

To proceed further we use the Peter--Weyl theorem: the collection of matrix elements of irreducible unitary representations of $G$ forms an orthonormal basis for
$L^2(G)$. The irreducible representations of $G$ are labelled by a pair of half integers $(I,J)$, interpreted as isospin and spin. Integer representations are even under the antipodal map, half-integer representations are odd. Hence, if $B$ is even (odd) then the wavefunction lies in the subspace spanned by $I$, $J$ both integer (non-integer). The $(I,J)$ subspace has dimension
$(2I+1)^2(2J+1)^2$ and is spanned by the functions
\[
\rho_I(g)_{ij}\rho_J(h)_{kl},\qquad 1\leq i,j\leq 2I+1,\quad 1\leq k,l\leq 2J+1,
\]
where $\rho_L\colon {\rm SU}(2)\ra {\rm GL}(2L+1,\C)$ is the spin $L$ representation of ${\rm SU}(2)$.
We can decompose this space further into subspaces with $(i,k)$ fixed. The Hamiltonian preserves each such subspace, and acts isomorphically on each (by left $G$-invariance of the metric and potential). Hence, we can, without loss of generality, fix $(I,J)$, set $i=k=1$, and consider the space spanned by
\[
\rho_I(g)_{1i}\rho_J(h)_{1j},\qquad 1\leq i\leq 2I+1,\quad 1\leq j\leq 2J+1.
\]
Let us denote these functions
\[
|i,j\rangle(g,h)=\rho_I(g)_{1i}\rho_J(h)_{1j}.
\]
Then our wavefunction (on $G\Gamma_j$) takes the form
\begin{gather}\label{ot}
\psi(s,g,h)=\sum_{i=1}^{2I+1}\sum_{j=1}^{2J+1}a_{ij}(s)|i,j\rangle(g,h),
\end{gather}
and represents a quantum state of isospin $I$ and spin $J$.

We may arrange the coefficients $a_{ij}(s)$ into a column vector indexed
$11,12,\ldots,1(2J+1),\allowbreak 21,22,\ldots,(2J+1),\ldots,(2I+1)1,(2I+1)2,\ldots,(2I+1)(2J+1)$,
call it $v(s)$ say. Then,
having fixed $(I,J)$, the differential operators $\theta_i$ act on
$v(s)$ by matrix multiplication according to the corresponding representations of $\mathfrak{su}(2)$,
\begin{gather}\label{scgifu}
\theta_i\equiv \theta_i^I\otimes \I_{2J+1},\qquad
\theta_{i+3}\equiv \I_{2I+1}\otimes \theta_i^J.
\end{gather}
For $I=J=1/2$, one has $\theta_i^{1/2}=-\frac{{\rm i}}{2}\tau_i$, for example. In this way, the Hamiltonian on $G\Gamma_j$ is reduced to a matrix-valued  ordinary differential operator acting on $v\colon[0,1]\ra\C^{(2I+1)(2J+1)}$, obtained explicitly by substituting~\eqref{scgifu} into~\eqref{Lap} and~\eqref{Ham}.

Suppose now that $(g_0,h_0)$ is a symmetry of $x(s)$ with compensating permflip
$(\sigma,\s)$, that is,
\begin{gather}\label{lauhur}
(g_0,h_0)\cdot x(s)=(\sigma,\s)\cdot x(s).
\end{gather}
Then, for all $(g,h)\in G$,
\begin{gather}\label{onte}
(g,h)\cdot((g_0,h_0)\cdot x(s))=(g,h)\cdot((\sigma,s)\cdot x(s))=
(\sigma,s)\cdot((g,h)\cdot x(s))
\end{gather}
(just act on equality~\eqref{lauhur} with the $G$ action and note that permflips commute with this action).
Note that the leftmost configuration in~\eqref{onte} is $(gg_0,hh_0)\cdot x(s)$.
Hence, the FR constraint demands that
\[
\psi(s,gg_0,hh_0)=(-1)^{\sgn(\sigma,\s)}\psi(s,g,h).
\]
Substituting our expansion for $\psi$,~\eqref{ot}
\begin{align*}
\sum a_{ij}(s)\rho_I(gg_0)_{1i}\rho_J(hh_0)_{1j}&=
\sum a_{ij}(s)\rho_I(g)_{1k}\rho_I(g_0)_{ki}\rho_J(h)_{1l}\rho_J(h_0)_{lj}\\
&=\sum \rho_I(g_0)_{ki}\rho_J(h_0)_{lj}a_{ij}(s)|k,l\rangle(g,h)\\
&=(-1)^{\sgn(\sigma,\s)}\sum a_{kl}(s)|k,l\rangle(g,h).
\end{align*}
Hence, our column vector $v(s)$ satisfies
\begin{equation}\label{sosm}
[\rho_I(g_0)\otimes \rho_J(h_0)]\: v(s)= (-1)^{\sgn(\sigma,\s)}v(s),
\end{equation}
where $\rho_I(g_0)\otimes \rho_J(h_0)$ is the tensor product of the matrices
$\rho_I(g_0)$ and $\rho_J(g_0)$.

In practice, we find that $x(s)$, $0<s<1$, the interpolating points on the energy minimizing curves $\Gamma_j$ have no nontrivial symmetries. The endpoints, that is, the local energy minimizers $x(0)$ and $x(1)$ frequently {\em do} have nontrivial symmetries, however. The FR constraints~\eqref{sosm} must then be incorporated into the boundary conditions imposed on $v(s)$. These boundary conditions must also be chosen to render $H$ self-adjoint. In the case where an endpoint is a
polyvalent vertex of $\Gamma$, one must also impose that $\psi$ is continuous at the vertex, and that probability flux is conserved
\cite{fabcsa}. In the sequel, we shall restrict attention to the simplest nontrivial scenario, where $\Gamma$ is a single edge connecting two distinct vertices, so that this latter requirement is not a~consideration. Even here, the FR constraints and self-adjointness
make the boundary conditions quite subtle.

\section[Quantizing the B=7 Skyrmion]{Quantizing the $\boldsymbol{B=7}$ Skyrmion}

For $B=7$, the graph $\Gamma$ has a single edge $x(s)$ connecting two energy minimizers denoted~7a and~7b, as seen in Table~\ref{tab:graphs}. The start, intermediate saddle point and end configurations are depicted in Figure~\ref{fig:B7configs}.
The 7a minimizer $x(0)$ has the symmetries of the following configuration:
\begin{center}
\begin{tabular}{|c|c|c|}\hline
label & position & orientation \\ \hline
1& $(0,0,-1)$ & $1$\\
2&$(0,0,1)$ & $1$\\
3&$(0,-1,0)$ & ${\rm i}$\\
4&$(0,1,0)$ & ${\rm i}$\\
5&$(-1,0,0)$ & ${\rm j}$\\
6&$(1,0,0)$ & ${\rm j}$\\
7&$(-1,-1,-1)$ & ${\rm k}$\\ \hline
\end{tabular}
\end{center}
To visualize this, think of particles 1 to 6 as being on the vertices of the octohedron where the coordinate axes intersect the unit sphere, and particle 7 as the extra particle hovering below one of the triangular faces.
Following~\cite{gilharkirmayspe}, we are specifying orientation by a unit quaternion using the identification
\[
q=q_0+q_1{\rm i}+q_2{\rm j}+q_3{\rm k} \leftrightarrow U=q_0\I_2-{\rm i}(q_1\tau_1+q_2\tau_2+q_3\tau_3).
\]

The 7a minimizer is symmetric under the action of $(g_0,h_0)$, where
\begin{gather*}
g_0=\frac12(1-{\rm i}-{\rm j}+k)\equiv\frac12(\I_2+{\rm i}(\tau_1+\tau_2-\tau_3)),\\
h_0=\frac12(1+{\rm i}+{\rm j}+{\rm k})\equiv\frac12(\I_2-{\rm i}(\tau_1+\tau_2+\tau_3)).
\end{gather*}
To see this, note that
\[
R(h_0)=\left(\begin{matrix}
0&0&1\\
1&0&0\\
0&1&0
\end{matrix}\right)
\]
which rotates by $120^\circ$ about the line containing the origin and $(-1,-1,-1)$.
This induces the permutation of positions
\[
1\mapsto3\mapsto5\mapsto1, \qquad
2\mapsto4\mapsto6\mapsto2,\qquad
7\mapsto7.
\]
Further, $q\mapsto h_0qg_0^{-1}$, acts on the unit quaternions as
\[
1\mapsto {\rm j}, \qquad
{\rm i}\mapsto -1, \qquad
{\rm j}\mapsto -{\rm i}, \qquad
{\rm k}\mapsto {\rm k}.
\]
So $(g_0,h_0)x(0)$ coincides with the image of $x(0)$ under the perm-flip
\[
((1,3,5)(2,4,6),+1,+1,-1,-1,-1,-1,1)\in P_7.
\]
This has sign $+1$ (the permutation is even, being a product of disjoint even cycles, i.e., cycles of odd length, and there are an even number of orientation sign flips).

It follows that the left endpoint value $v(0)$ of the wavefunction must lie in the $+1$ eigenspace of the matrix
$\rho_I(g_0)\otimes\rho_J(h_0)$. We will always choose $I=1/2$ \big(corresponding to Li${}^7$ or Be${}^7$\big), but will consider the spin values $J=1/2,3/2,5/2$ and $7/2$.
For $I=J=1/2$, $\rho_I=\rho_J=\id$, so
\[
\rho_{1/2}(g_0)\otimes\rho_{1/2}(h_0)=g_0\otimes h_0=\frac12\left(\begin{matrix}
-{\rm i}&-1&1&-{\rm i}\\
-{\rm i}&\hphantom{-}1&1&\hphantom{-}{\rm i}\\
\hphantom{-}{\rm i}&\hphantom{-}1&1&-{\rm i}\\
-{\rm i}&-1&1&\hphantom{-}{\rm i}
\end{matrix}\right),
\]
whose $+1$ eigenspace $F_0^{1/2}$ is spanned by
\[
\left(\begin{matrix}
2\\-(1+{\rm i})\\1+{\rm i}\\2{\rm i}\end{matrix}\right),\qquad
\left(\begin{matrix}0\\1\\1\\0\end{matrix}\right).
\]
For higher spin, we must construct the spin $J$ representation of
$h_0$, a routine calculation that we relegate to Appendix \ref{appendix2}, then compute the $+1$ eigenspace $F_0^{J}$ of
$\rho_{1/2}(g_0)\otimes\rho_J(h_0)$.
To ensure self-adjointness of $H$, we demand that
\begin{gather}\label{bc0}
v(0)\in F_0^{J},\qquad \mbox{and}\qquad v'(0)\in \big(F_0^{J}\big)^\perp.
\end{gather}

The 7b minimizer $x(1)$ has the symmetries of the following configuration:
\begin{center}
\begin{tabular}{|c|c|c|}\hline
label & position & orientation \\ \hline
1& $(1,0,0)$ & $-{\rm j}$\\
2&$(0,0,-1)$ & $\hphantom{-}1$\\
3&$(-1,0,0)$ & $-{\rm j}$\\
4&$(0,0,1)$ & $\hphantom{-}1$\\
5&$(1,1,1)$ & $-{\rm k}$\\
6&$(-1,1,-1)$ & $\hphantom{-}{\rm k}$\\
7&$(0,1,0)$ & $\hphantom{-}{\rm i}$\\ \hline
\end{tabular}
\end{center}
To visualize this, think of particles 1 to 4 as on the vertices of the unit square in the plane $y=0$ (where the unit circle intersects the $x$ and $z$ axes) and particles 5 to 7 as lying on the diagonal line $x=z$ in the plane $y=1$.

This is symmetric under the action of $(g_1,h_1)$, where
\begin{gather*}
g_1 = {\rm j}\equiv-{\rm i}\tau_2,\qquad
h_1 = -{\rm j}\equiv {\rm i}\tau_2.
\end{gather*}
To see this, note that
\[
R(h_1)=\left(\begin{matrix}
-1&0&\hphantom{-}0\\
\hphantom{-}0&1&\hphantom{-}0\\
\hphantom{-}0&0&-1
\end{matrix}\right),
\]
which rotates by $180^\circ$ about the $y$ axis.
This induces the permutation of positions
\[
1\mapsto3\mapsto1, \qquad
2\mapsto4\mapsto2,\qquad
5\mapsto6\mapsto5,\qquad
7\mapsto7.
\]
Further, $q\mapsto h_1qg_1^{-1}$, acts on the unit quaternions as
\[
1\mapsto -1, \qquad
{\rm i}\mapsto {\rm i}, \qquad
{\rm j}\mapsto -{\rm j}, \qquad
{\rm k}\mapsto {\rm k}.
\]
So $(g_1,h_1)x(1)$ coincides with the image of $x(1)$ under the perm-flip
\[
((1,3)(2,4)(5,6),-1,-1,-1,-1,-1,-1,1)\in P_7.
\]
This has sign $-1$ since the permutation is odd (being a product of 3 transpositions) and the there are an even number of orientation flips.
It follows that the right endpoint value $v(1)$ of the wavefunction must lie in the $-1$ eigenspace of the matrix
$\rho_I(g_1)\otimes\rho_J(h_1)$ which, for $I=1/2$, we denote $F_1^J$.
For $J=1/2$, we have
\[
\rho_{1/2}(g_1)\otimes\rho_{1/2}(h_1)=g_1\otimes h_1=\left(\begin{matrix}
\hphantom{-}0&0&0&-1\\
\hphantom{-}0&0&1&\hphantom{-}0\\
\hphantom{-}0&1&0&\hphantom{-}0\\
-1&0&0&\hphantom{-}0
\end{matrix}\right),
\]
whose $-1$ eigenspace $F_1^{1/2}$ is spanned by
\[
\left(\begin{matrix}1\\0\\0\\1\end{matrix}\right),\qquad
\left(\begin{matrix}\hphantom{-}0\\ \hphantom{-}1\\-1\\ \hphantom{-}0\end{matrix}\right).
\]
To ensure self-adjointness of $H$, we demand that
\begin{gather}\label{bc1}
v(1)\in F_1^{J},\qquad \mbox{and}\qquad v'(1)\in \big(F_1^{J}\big)^\perp.
\end{gather}
The dimensions of the allowed spaces $F_{0,1}^J$ and their orthogonal complements for $J=1/2$, $3/2$, $5/2$ and $7/2$ are summarized in Table~\ref{table FRspaces}.

\begin{table}[t]\centering
\begin{tabular}{|c|cccc|}\hline
spin $J$ & $\dim F_0^J$ & $\big(\dim F_0^J\big)^\perp$ & $\dim F_1^J$ & $\big(\dim F_1^J\big)^\perp$ \\ \hline
1/2 & 2 & 2 & 2 & 2 \\
3/2 & 2 & 6 & 4 & 4 \\
5/2 & 4  & 8  & 6  & 6  \\
7/2 & 6  & 10  & 8  & 8  \\ \hline
\end{tabular}
\caption{The dimensions of the symmetry compatible state spaces at the
endpoints of the energy minimizing path between the 7a and 7b minimizers, for isospin 1/2.}
\label{table FRspaces}
\end{table}

Having fixed $I=1/2$ and $J$, our wave function is a map
$v\colon[0,1]\ra\C^{4J+2}$ satisfying the boundary conditions
\eqref{bc0} and~\eqref{bc1}. To find the ground state in the spin $J$ sector, we solve the minimization problem for the energy functional
\[
E(v)=\ip{v,Hv}_{L^2}
\]
over all $v$ satisfying the boundary conditions with $\|v\|_{L^2}=1$ (where, in computing the $L^2$ inner product and norm, we equip $[0,1]$ with the measure $\sqrt{|\Lambda(s)|q(s)}$). In practice, we achieve this by constrained gradient flow (gradient flow for $E$ tangent to the constraints $\|v\|_{L^2}=1$ and the boundary conditions).

To proceed further we must calibrate the model, that is, choose energy and length scales by fitting to experimental data. We present here results for the calibration proposed in~\cite{gilharspe}, chosen so that the quantized $I=J=1/2$, $B=1$ soliton has the energy and charge radius of the nucleon.
This has energy and length units
\[
E_*=10.72~\text{MeV},\qquad l_*=0.6505~\text{fm},
\]
and implies that $\hbar = 28.2752$ in these units~\cite{gilharkirmayspe}. One expects the kinetic term in $H$ to dominate in this regime (recall that $V(s)$ varies very little over the graph $\Gamma$, see Table~\ref{tab:graphs}) and, indeed, we find the results do not change significantly if the potential is omitted entirely.

\begin{table}[t]\centering
\begin{tabular}{|c|c|c|c|}\hline
spin $J$ & $E$ & $E/B$ & $E/E_{7a}$ \\ \hline
1/2 & 6948.63 &992.661  & 1.00096  \\ 
3/2 & 6950.02 & 992.860 & 1.00116  \\ 
5/2 & 6951.73 & 993.105 & 1.00141    \\ 
7/2 & 6953.88 & 993.411 & 1.00171  \\ \hline 
\end{tabular}
\caption{The energies of the wavefunctions presented in Figure~\ref{fig:wavefn} for the energy minimizing path between the 7a and 7b minimizers, for spins $1/2$ to $7/2$ in MeV and normalised by the energy of the 7a minimizer.}
\label{tab:waveEn}
\end{table}

\begin{figure}[t]\centering
\includegraphics[width=0.5\linewidth]{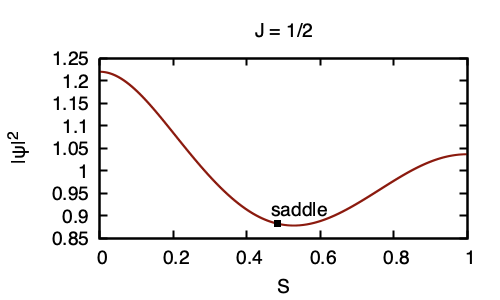}\includegraphics[width=0.5\linewidth]{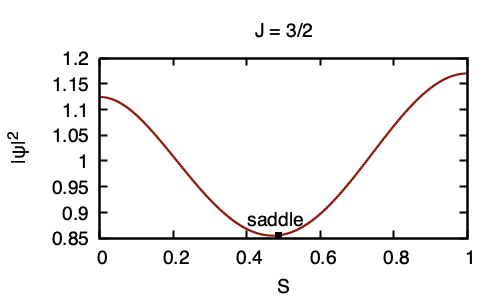}
\includegraphics[width=0.5\linewidth]{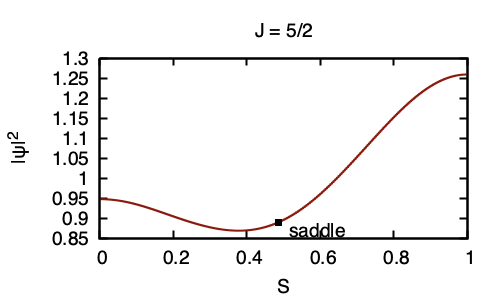}\includegraphics[width=0.5\linewidth]{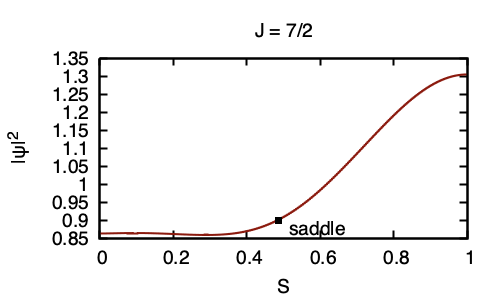}

\caption{Plots of the normalized probability densities $|\psi(S)|^2$ for the $B=7$ graph, as a function of the fractional volume coordinate $S$ defined in equation~\eqref{peso}, for spins $J = 1/2$ to $J = 7/2$ and isospin $I = 1/2$. The global energy minimising configuration ($a$) is at $S = 0$ and a local energy minimizer ($b$) at $S=1$. The saddle point on the curve joining them occurs at $S=0.485$.  The graph and energies are given in Table~\ref{tab:graphs}.}
\label{fig:wavefn}
\end{figure}

The resulting wavefunctions for $J=1/2$ to $7/2$ are plotted in Figure~\ref{fig:wavefn}. It is useful to change coordinate from $s\in[0,1]$ to $S\in[0,1]$ defined as follows. Let
\[
{\rm Vol}(s_0,s_1)=\int_{[s_0,s_1]\times G}\vol,
\]
the volume of the slice of $G\Gamma$ between $s=s_0$ and $s=s_1$, and
\begin{gather}\label{peso}
S(s)=\frac{{\rm Vol}(0,s)}{{\rm Vol}(0,1)},
\end{gather}
the fraction of the total volume of $G\Gamma$ to the ``left'' of $s$. Note that the $7a$ and $7b$ minima are still at $S=0$ and $S=1$. The advantage of this coordinate is that the induced measure on the interval $[0,1]$ is simply ${\rm d}S$, so equal $S$ intervals represent chunks of $G\Gamma$ of equal volume, making it easier to visualize the quantum states' localization properties. In this coordinate, the saddle point occurs at $S=0.485$.

We note that the wavefunction is far from localized around the lower energy minimizer.  In fact as $J$ increases the wavefunction becomes more localized around the higher energy minimizer.
Table \ref{table FRspaces} suggests a possible qualitative explanation for this: for higher spin, the FR constraints are less restrictive at the $b$ end than at the $a$ end and so, since quantum mechanics exacts an energetic cost for confinement, the state tends to avoid the more tightly constrained endpoint. In all cases, the state spreads significantly over the whole moduli space.
This suggests that rigid body quantization is a poor approximation to quantizing over the complete configuration space.  The numerical energies are given in Table \ref{tab:waveEn}, and should be compared with the experimental mass of Li${}^7$, which is 6535 MeV.  Note that energy increases monotonically with spin $J$, in conflict with experimental data which order the states $3/2$, $1/2$, $7/2$, $5/2$. This suggests that the lightly bound model is unlikely to give a good account of atomic nuclei.

\appendix

\section{Appendix: numerical minimizers}
\label{appendix1}

The table below records the interaction energies ($V_{\rm pp}-BM$) of the approximate local minimizers of $V_{\rm pp}$ constructed in~\cite{gilharkirmayspe}  by simulated annealing ($E_{\rm ann}$), compared with their interaction energies after relaxation by gradient flow ($E_{\rm grad}$).  The final column is a measure of the distance between the initial $X^{\rm ann}$ and final $X^{\rm grad}$ configurations, each of which we may think of as a point in \smash{$\big(\R^3\times S^3\big)^B\subset\big(\R^{7}\big)^{B}$}. We measure distance in this space using the norm
\[
\|X\|=\max\{|X_i|_{\R^7}\colon i=1,2,\ldots,B\},
\]
where $|\cdot|_{\R^7}$ is the Euclidean norm on $\R^7$. Hence $\big\|X^{\rm ann}-X^{\rm grad}\big|$ is the maximum distance (in position-orientation space) moved by any of the $B$ particles after relaxation.

  The relaxed minimizers were used as endpoint data for the NEB method.  The
  solutions are labelled in energy order and those that have changed label due to the more accurate results have been marked with a $\dagger$.  Additional local minimizers found while using the NEB method are marked with a *. Note that one claimed minimizer for degree 9 (labelled b in~\cite{gilharkirmayspe}) was not a~true minimizer and reduced to the configuration c in that paper and has been removed.

\begin{table}[t]\centering
\begin{tabular}{|c|c|c|c|c|c|}\hline
$B$ & label &  $E_{\rm ann}$ & $E_{\rm grad}$ & $E_{\rm ann} - E_{\rm grad}$  & $\|X^{\rm ann}-X^{\rm grad}\|$ \\ \hline
3 & a & $-0.929766$ & $-0.930915$ & 0.0011488 & 0.0166381 \\ \hline
4 & a & $-1.8618$ & $-1.86183$ & 3.4363 $\times 10^{-5}$ & 0.00526994 \\ \hline
\multirow{2}{*}{5} & a & $-2.32215$ & $-2.33824$ & 0.0160941\hphantom{0} & 0.0407176 \\
 & b & $-2.18172$ & $-2.18481$ & 0.00308473 & 0.0180328 \\ \hline
\multirow{3}{*}{6} & a & $-3.22892$ & $-3.22909$ & 0.000168902 & 0.00836376\\
 & b & $-3.10659$ & $-3.11693$ & 0.0103361\hphantom{00} & 0.0235825\hphantom{0} \\
 & c & $-3.03345$ & $-3.04656$ & 0.0131102\hphantom{00} & 0.0421782\hphantom{0} \\ \hline
\multirow{2}{*}{7} & a & $-4.05417$ & $-4.05734$ & 0.00316932 & 0.0140242 \\
 & b & $-3.84305$ & $-3.8953$ & 0.0522464\hphantom{0} & 0.0717353 \\ \hline
\multirow{3}{*}{8} & a & $-4.88421$ & $-4.88901$ & 0.00479826 & 0.0182521 \\
 & b & $-4.86001$ & $-4.86923$ & 0.00922791 & 0.0237361 \\
 & c & $-4.74428$ & $-4.78119$ & 0.0369082\hphantom{0} & 0.0502746 \\ \hline
\multirow{5}{*}{9} & a & $-5.66335$ & $-5.66437$ & 0.00102358 & 0.0148732 \\
 & b$^\dagger$ & $-5.57431$ & $-5.59792$ & 0.0236121\hphantom{0} & 0.0447026\\
 & c$^\dagger$ & $-5.42859$ & $-5.48298$ & 0.0543869\hphantom{0} & 0.0762035 \\
 & d$^\dagger$ & $-5.42974$ & $-5.4595$ & 0.0297558\hphantom{0} & 0.0497296 \\ \hline
\multirow{6}{*}{10} & a$^\dagger$ & $-6.44096$ & $-6.44306$ & 0.00209978\hphantom{0} & 0.0155593\\
& b$^\dagger$ & $-6.44139$ & $-6.44207$ &0.000672919 & 0.0154335 \\
& c$^\dagger$ & $-6.22549$ & $-6.30734$ & 0.0818569\hphantom{00} & 0.0729622\\
 & d$^\dagger$ & $-6.22416$ & $-6.28447$ & 0.060308\hphantom{000} & 0.0773449 \\
 & e$^\dagger$ & $-6.24819$ & $-6.27703$ & 0.028844\hphantom{000} & 0.0493404\\
 & f & $-6.06189$ & $-6.19396$ & 0.132066\hphantom{000} & 0.07183\hphantom{0}\\
 & g* & --& $-6.13281$ & -- & -- \\
 & h* & -- & $-6.1205\hphantom{0}$ & -- & -- \\ \hline
\end{tabular}
\caption{The numerical minimizers found by simulated annealing in~\cite{gilharkirmayspe} before and after relaxation by energy gradient flow.}
\end{table}

\section[Appendix: higher spin representations of SU(2)]{Appendix: higher spin representations of $\boldsymbol{{\rm SU}(2)}$}\label{appendix2}

We record here an elegant construction of the higher spin representations of
${\rm SU}(2)$ which, while certainly not new, does not have (as far as we are aware) a simple description in the literature. The account below is based partially on unpublished notes of Urbanik~\cite{urb}.

Let $n=2J$ and $V_n$ be the set of homogeneous polynomials in two complex variables $(z_1,z_2)$ of degree $n$, with complex coefficients. This is an $(n+1)$-dimensional complex vector space spanned by
\begin{gather}\label{monomials}
z_1^n, \ z_1^{n-1}z_2, \ z_1^{n-2}z_2^2,\ \ldots, \ z_2^n.
\end{gather}
Define a left action of ${\rm SU}(2)$ on $V_n$ by
$
A\colon   f\mapsto f\circ A^\dagger$.
That is, associate to the ${\rm SU}(2)$ matrix
\[
A=\left(\begin{matrix}\hphantom{-}\alpha & \beta \\ -\bar\beta & \bar\alpha
\end{matrix}\right), \qquad |\alpha|^2+|\beta|^2=1,
\]
the map $V_n\ra V_n$ which takes a homogeneous polynomial $f(z_1,z_2)$ to the
homogeneous polynomial
\begin{equation}\label{su2action}
f_A(z_1,z_2)=f\big(\bar\alpha z_1-\beta z_2,\bar\beta z_1 +\alpha z_2\big).
\end{equation}
Note that $(f_A)_{A'}=f_{AA'}$, so this is indeed a representation of
${\rm SU}(2)$ on $V_n$, which can be shown to be irreducible~\cite{urb}.

To make the representation unitary, as we require, we must equip $V_n$ with an ${\rm SU}(2)$ invariant hermitian inner product. The naive choice, wherein the monomials
in~\eqref{monomials} are declared orthonormal does not work. Instead, we identify $V_n$ with $H^0(\mathcal{O}(n))$, the space of holomorphic sections of the degree $n$ holomorphic line bundle $\mathcal{O}(n)$ over $\CP^1$, as follows.
 Think of $(z_1,z_2)$ as a point in the fibre $L_p$ of the tautological bundle
${\mathcal O}(-1)$ over the point $p=[z_1,z_2]\in \CP^1$. Then the map $(z_1,z_2)\mapsto z_1^{n-k}z_2^k$ is an $n$-fold multilinear map $L_p\times L_p\times \cdots\times L_p\ra\C$, that is, an element of $L_p^*\otimes L_p^*\otimes\cdots\otimes L_p^*$, the fibre of ${\mathcal O}(n)$ over $p$. So we can get an inner product on~$V_n$ by identifying it with $H^0({\mathcal O}(n))$ and putting an inner product on that.

There is a natural ${\rm SU}(2)$ invariant fibre metric on ${\cal O}(n)$, which equips $H^0({\mathcal O}(n))$ with an $L^2$ inner product. If we trivialize ${\mathcal O}(n)$ over the open set $z_2=0$, we can specify the value of the section at $[z,1]$ by saying what it does to $(z,1)$. This amounts to identifying the homogeneous polynomial $f(z_1,z_2)$ with
\[
f(z)=f(z,1)=a_0 z^n+a_1z^{n-1}+\cdots+a_n.
\]
Then the $L^2$ inner product of two sections is
\[
\ip{f,g}_{L^2}=4\int_{\C}\frac{\ol{f(z)}g(z)}{(1+|z|^2)^{n+2}}\,{\rm d}x {\rm d}y.
\]
It is easy to see that the monomials $z^j$, $z^k$ are orthogonal with respect to this inner product if $j\neq k$, but they do not all have the same length. To get a unitary representation of ${\rm SU}(2)$ on~$V_n$, we should compute its matrix with respect to the orthonormal basis
\[
\frac{z^{n-k}}{\|z^{n-k}\|_{L^2}},\qquad k=0,1,\ldots,n.
\]

To illustrate, let us compute the spin $J=3/2$ representation of the
rotational symmetry of the $7b$ solution,
\[
h_1=\left(\begin{matrix}
\hphantom{-}0 & 1 \\ -1 & 0
\end{matrix}\right).
\]
We must compute the action of $h_1$ on $V_3\equiv H^0(\mathcal{O}(3))$, which is spanned by the monomials $z^3$,~$z^2$,~$z$ and~$1$. It is easy to check that
\[
\big\|z^3\big\|_{L^2}^2=\pi,\qquad
\big\|z^2\big\|_{L^2}^2=\frac\pi3,\qquad
\big\|z^1\big\|_{L^2}^2=\frac\pi3,\qquad
\big\|1\big\|_{L^2}^2=\pi.
\]
Hence, an orthonormal basis for $V_3$ is provided by the homogeneous polynomials
\[
\frac{z_1^3}{\sqrt\pi},\qquad
\sqrt{\frac{3}{\pi}}z_1^2z_2,\qquad
\sqrt{\frac{3}{\pi}}z_1z_2^2,\qquad
\frac{z_2^3}{\sqrt\pi},
\]
whose images under $h_1$, using the action~\eqref{su2action}, are
\begin{gather*}
\frac{z_1^3}{\sqrt\pi} \mapsto -\frac{z_2^3}{\sqrt\pi},\qquad
\sqrt{\frac{3}{\pi}}z_1^2z_2 \mapsto \sqrt{\frac{3}{\pi}}z_1z_2^2,\qquad
\sqrt{\frac{3}{\pi}}z_1z_2^2 \mapsto -\sqrt{\frac{3}{\pi}}z_1^2z_2,\qquad
\frac{z_2^3}{\sqrt\pi} \mapsto \frac{z_1^3}{\sqrt\pi}.
\end{gather*}
Hence the spin $3/2$ representation of $h_1$ is
\[
\rho_{3/2}(h_1)=\left(\begin{matrix}
0&\hphantom{-}0&0&-1\\
0&\hphantom{-}0&1&\hphantom{-}0\\
0&-1&0&\hphantom{-}0\\
1&\hphantom{-}0&0&\hphantom{-}0
\end{matrix}\right).
\]

To compute the spin $J$ representation of $\theta_i$, the generators of $\mathfrak{su}(2)$, we apply the above procedure to generating curves
\[
A_i(t)=\cos\frac{t}{2}\I_2-{\rm i}\sin\frac{t}{2}\tau_i
\]
for $\theta_i$ and differentiate at $t=0$. For $J=3/2$, one finds, for
example,
\begin{gather*}
\theta_1^{3/2}=\d\rho_{3/2}(\theta_1)=
\left(\begin{matrix}
0&\frac{\sqrt{3}}{2}{\rm i}&0&0\\
\frac{\sqrt{3}}{2}{\rm i}&0&{\rm i}&0\\
0&i&0&\frac{\sqrt{3}}{2}{\rm i}\\
0&0&\frac{\sqrt{3}}{2}{\rm i}&0
\end{matrix}\right),\\
\theta_2^{3/2}=\d\rho_{3/2}(\theta_2)=
\left(\begin{matrix}
0&-\frac{\sqrt{3}}{2}&0&0\\
\frac{\sqrt{3}}{2}&0&-1&0\\
0&1&0&-\frac{\sqrt{3}}{2}\\
0&0&\frac{\sqrt{3}}{2}&0
\end{matrix}\right),\nonumber \\
\theta_3^{3/2}=\d\rho_{3/2}(\theta_3)=
\left(\begin{matrix}
\frac32{\rm i}&0&0&0\\
0&\frac12{\rm i}&0&0\\
0&0&-\frac12{\rm i}&0\\
0&0&0&-\frac32{\rm i}\end{matrix}\right).
\end{gather*}
These calculations are easily automated.

\subsection*{Acknowledgements}

We would like to thank Chris Halcrow and Derek Harland for useful conversations. This work was partially funded by
the UK Engineering and Physical Sciences ResearchCouncil, through grant
EP/P024688/1.


\pdfbookmark[1]{References}{ref}

\LastPageEnding

\end{document}